\documentclass[conference]{IEEEtran}

\usepackage[normalem]{ulem}

\usepackage[noend]{algpseudocode}
\usepackage{algorithm}
\usepackage{booktabs}
\usepackage{amssymb}
\usepackage{changepage}
\usepackage{pifont}

\usepackage{fancyhdr}
\usepackage[normalem]{ulem}
\usepackage{microtype}
\usepackage{enumitem}

\makeatletter
\@ifpackageloaded{subfig}
{}
{\usepackage[labelformat=simple,subrefformat=parens,caption=false]{subfig}
 
 }
\makeatother
\usepackage{graphicx}

\usepackage{color}
\usepackage{listings}
\usepackage{relsize}
\usepackage{multirow}
\usepackage[normalem]{ulem} 

\usepackage{xspace}

\newcommand{\originalgrumbler}[2]{\begin{quote}\textcolor{blue}{\sl{\bf #1 says:} #2}\end{quote}}
\newcommand{\grumbler}[2]{\originalgrumbler{#1}{#2}}

\newcommand{\mike}[1]{\grumbler{Mike}{#1}}

\newcommand{\swarnendu}[1]{\grumbler{Swarnendu}{#1}}

\newcommand{\brandon}[1]{\grumbler{Brandon}{#1}}

\newcommand{\rui}[1]{\grumbler{Rui}{#1}}

\newcommand{\vignesh}[1]{\grumbler{Vignesh}{#1}}

\definecolor{darkgreen}{rgb}{0,0.4,0}

\newcommand{\sfsmaller}{}

\newcommand{\bench}[1]{\textsf{\sfsmaller#1}}
\newcommand{\code}[1]{\textsf{\sfsmaller#1}}

\newcommand{\eg}{e.g.\xspace}
\newcommand{\ie}{i.e.\xspace}

\newcommand{\etal}{et al.\xspace}

\makeatletter
\@ifundefined{state}{%
}{}
\makeatother

\usepackage{etoolbox}

\newcommand\later[1]{\begin{quote}\textcolor{darkgreen}{\textbackslash \textbf{later\{}} #1 \textcolor{darkgreen}{\}}\end{quote}}
\renewcommand{\later}[1]{}

\newcommand\notes[1]{\begin{quote}\textcolor{darkgreen}{\textbackslash \textbf{notes\{}} #1 \textcolor{darkgreen}{\}}\end{quote}}
\renewcommand{\notes}[1]{}

\newcommand{\barc}{Neat\xspace}
\newcommand{\Barc}{\barc}

\newcommand{\neat}{\barc}
\newcommand{\Neat}{\Barc}

\newcommand{\mesi}{MESI\xspace}
\newcommand{\Mesi}{MESI\xspace}

\newcommand{\sarc}{SARC\xspace}
\newcommand{\Sarc}{SARC\xspace}

\newcommand{\wb}{write-back\xspace}
\newcommand{\wbs}{write-backs\xspace}
\newcommand{\Wb}{Write-back\xspace}

\newcommand{\si}{self-in\-va\-li\-da\-tion\xspace}
\newcommand{\sis}{self-in\-va\-li\-da\-tions\xspace}
\newcommand{\Si}{Self-in\-va\-li\-da\-tion\xspace}

\newcommand{\cmt}{commit\xspace}
\newcommand{\Cmt}{Commit\xspace}

\newcommand{\mesicfg}{\code{MESI}\xspace}

\newcommand{\bcunopt}{\code{Neat base}\xspace}
\newcommand{\bcpi}{\code{Neat pi-only}\xspace}
\newcommand{\bcbf}{\code{Neat}\xspace}
\newcommand{\bcopt}{\bcbf}

\newcommand{\ntunopt}{\bcunopt}
\newcommand{\ntpi}{\bcpi}
\newcommand{\ntbf}{\bcbf}
\newcommand{\ntopt}{\ntbf}

\newcommand{\ntpsrw}{\code{Neat cla}\xspace}
\newcommand{\ntcla}{\ntpsrw}

\newcommand{\vipsunopt}{\code{VIPS unopt}\xspace}

\newcommand{\vipspsrw}{\code{VIPS cla}\xspace}
\newcommand{\vipscla}{\vipspsrw}

\newcommand{\sarcws}{\textcolor{red}{\code{S/N hybrid}}\xspace}
\newcommand{\scunopt}{\code{SARC}\xspace}

\newcommand{\Scunopt}{\scunopt}

\newcommand{\ws}{write signature\xspace}

\newcommand{\wss}{write signatures\xspace}

\newcommand{\opt}{optimization\xspace}

\newcommand{\cfg}{configuration\xspace}
\newcommand{\cfgs}{configurations\xspace}

\newcommand{\ack}{acknowledgment\xspace}

\newcommand{\vips}{VIPS\xspace}
\newcommand{\Vips}{VIPS\xspace}

\renewcommand\sfsmaller{\small}

\title{\Barc: Low-Complexity, Efficient
On-Chip Cache Coherence}

\makeatletter
\newcommand{\linebreakand}{%
  \end{@IEEEauthorhalign}
  \hfill\mbox{}\par
  \mbox{}\hfill\begin{@IEEEauthorhalign}
}
\makeatother

\author{\IEEEauthorblockN{Rui Zhang}
  \IEEEauthorblockA{
    \textit{Ohio State University}\\
    Columbus, OH, USA \\
    zhang.5944@osu.edu}
  \and
  \IEEEauthorblockN{Swarnendu Biswas}
  \IEEEauthorblockA{
\textit{Indian Institute of Technology}\\
    Kanpur, India \\
    swarnendu@cse.iitk.ac.in}
  \and
  \IEEEauthorblockN{Vignesh Balaji}
  \IEEEauthorblockA{
    \textit{Carnegie Mellon University}\\
    Pittsburgh, PA, USA \\
    vigneshb@andrew.cmu.edu}
\linebreakand
  \IEEEauthorblockN{Michael D. Bond}
  \IEEEauthorblockA{
    \textit{Ohio State University}\\
    Columbus, OH, USA    \\
    mikebond@cse.ohio-state.edu}
  \and
  \IEEEauthorblockN{Brandon Lucia}
  \IEEEauthorblockA{
    \textit{Carnegie Mellon University}\\
    Pittsburgh, PA, USA \\
    blucia@andrew.cmu.edu}
}

\begin{document}

\maketitle

\begin{abstract}

Cache coherence protocols such as \mesi that use writer-initiated invalidation
have high complexity---and sometimes have poor performance and energy usage,
especially under false sharing.
Such protocols require numerous transient states, a shared directory, and support for core-to-core communication, while also suffering under false sharing.
An alternative to MESI's writer-initiated invalidation is \emph{\si}, which
achieves lower complexity than MESI but
adds high performance costs or relies on programmer annotations or specific data access patterns.
\notes{or incur some of MESI's complexity.}%

This paper presents \emph{\neat}, a low-complexity, efficient cache coherence protocol.
\Neat uses \si,
thus avoiding \mesi's transient states, directory, and core-to-core communication requirements.
\Neat uses novel mechanisms that effectively avoid many unnecessary \sis.
An evaluation shows that \neat is simple and has lower verification complexity
than the \mesi protocol.
\Neat not only outperforms state-of-the-art \si protocols,
but its performance and energy consumption are comparable to
\mesi's, and it outperforms MESI under false sharing.

\end{abstract}

\notes{
\rui{Things done before the ISCA20 submission:
\\
1) Implement VIPS in two configs, \ie, w/ and w/o shared/private line classification.
\\
2) To solve the performance issue with several benchmarks, 
apply the special lock protocol to atomic accesses rather than treating them as region boundaries and 
\\
3) Merging important fixes from ARC/CE, \ie, 
a) including off-chip traffic into on-chip traffic.
and 
b) fixing the post-exec computation of mem accesses (reads and writes) in the exp framework.}
}%
\notes{
\rui{The ISCA20 reviewers were mainly concerned about the novelty of \neat, compared to prior work, especially the Denovo line of work.\\
If we consider the baseline DeNovo discussed in~\cite{denovo-gpu} which doesn't require programmer annotated regions to minimize self-invalidation overhead, DeNovo is merely "self-invalidation + registration", while registration looks much like SARC's writer coherence.}

\rui{From the reviews, it's also interesting to think about extending \neat to heterogeneous architectures or to be combined with hierarchical protocols.\\
In fact, the baseline \neat seems to use pretty much the same idea as the GPU+DRF coherence discussed in~\cite{denovo-gpu}, which is "self-invalidations + flushes/bulk writebacks". Similar to GPU+DRF, the major issue of using \neat for GPU coherence is unnecessary coherence actions at local synchronizations. DeNovo suffers the same issue too, but to a lesser degree, as its registration should be cheaper than \neat's bulk writebacks.
}
}%

\section{Introduction}
\label{sec:intro}


\later{\swarnendu{It is not super important, but a new edition of the primer book is available.}}%

Today's general-purpose processors have multiple cores with private and shared caches.
To provide the abstraction of shared memory in this context,
processors implement \emph{cache coherence}, which is defined by two invariants---the \emph{single writer, multiple readers} (SWMR) invariant and the \emph{data-value} invariant~\cite{coherence-primer}. 
The SWMR invariant requires that at any given time, a memory location can only be written by a single core or read by multiple cores.
The data-value invariant requires that a read must see the \emph{up-to-date} value of the corresponding memory location it reads.
Most multicore processors use a variant of the MESI cache coherence protocol~\cite{coherence-primer,illinois-mesi}.
MESI enforces the two invariants using \emph{writer-initiated invalidation} and \emph{ownership tracking}:
whenever a core writes to a cache line,
the protocol invalidates shared copies of the line, by tracking where in the system a valid copy or copies of a line reside;
the protocol also records the ID of the writer (\ie, owner ID) so that subsequent reads can be directed to the writer to get up-to-date values of the line.
As a result, MESI and its variants are efficient because they perform coherence actions only when accesses to a line by multiple cores conflict.

While often efficient, MESI has some serious drawbacks.
Protocol races---which occur even in executions of data-race-free programs---necessitate
transient states that complicate implementation and verification.
Optimized MESI implementations rely on a shared directory that maintains the coherence state and the owner ID of each line, typically containing an entry for every line in the shared cache.
MESI messages between cores' private caches and the shared cache require acknowledgments to ensure the SWMR and the data-value invariants,
\eg, for writer-initiated invalidations and dirty write-backs,
incurring latency.
MESI maintains coherence states at the granularity of cache lines and is consequently susceptible to false sharing~\cite{huron}.

Researchers have introduced new cache coherence designs that aim to be simpler than
MESI~\cite{sarc-coherence,vips-directoryless-noc-coherence,denovo,denovond,denovosync,
denovo-gpu}.
A key aspect of these designs is that they exploit the \emph{data-race-free} (DRF) assumption
of language-level memory consistency models~\cite{memory-models-cacm-2010}.
These simpler coherence protocols exploit DRF to
leverage the semantics
of synchronization to enforce coherence.
As a result, they do not require cores to
exchange eager invalidations that directly implement the SWMR invariant.
Using these techniques, a core instead {\em self-invalidates} its valid lines at
a synchronization acquire operation.
Further, by exploiting DRF, the coherence protocols do not require cores to write back dirty data or register the owner ID in the shared cache immediately upon each individual write.
A core can defer flushing dirty data or ownership registration to the shared cache until a synchronization release operation.
Assuming a DRF program, such self-invalidations and deferred flushes are sufficient to ensure coherence.

\begin{table*}
\begin{tabular}{l|l|l|l}
\multirow{2}{*}{Coherence protocol}       & Performance/energy & Performance/energy & \multirow{2}{*}{Complexity} \\
                                          & cost (w/o false sharing) & cost under false sharing & \\\hline
MESI~\cite{illinois-mesi} and its variants& Low                & High                & High \\
Prior \si protocols~\cite{sarc-coherence,vips-directoryless-noc-coherence,denovo,denovond,denovosync,denovo-gpu}
& Medium             & Low              & Low \\
\Neat (this paper)                        & Low--medium        & Low         & Low
\end{tabular}
\vspace*{0.5em}
\caption{Qualitative comparison of coherence protocols.
Note that \si protocols that retain some MESI features~\cite{sarc-coherence,denovosync}
incur a combination of the advantages and disadvantages of both.}
\label{tab:qualitative-comparison}
\end{table*}

\Si has \emph{potential} advantages over MESI:
lower protocol complexity (by avoiding MESI's numerous transient states),
lower power (mainly by eliminating MESI's coherence directory states), and
lower per-operation latency (by eliminating protocol acknowledgments, as well as cache invalidations due to false sharing).
At the same time, self-invalidation can degrade performance by
invalidating up-to-date lines, causing unnecessary cache misses.
Some self-invalidation protocols try to improve performance by
relying on mechanisms to infer data access patterns or on programmers to write in new languages or use annotations about which cache lines need to be invalidated~\cite{vips-directoryless-noc-coherence,denovo,denovond,denovo-gpu}.
To implement the data-value invariant,
some self-invalidation-based approaches
retain some of MESI's directory and protocol complexity~\cite{sarc-coherence};
or defer ownership registrations and dirty write-throughs by \emph{buffering} them,
which incur performance and energy cost when the buffers overflow~\cite{vips-directoryless-noc-coherence,denovo,denovond,denovo-gpu}.
Section~\ref{sec:background} details closely related prior approaches.

Table~\ref{tab:qualitative-comparison}'s first two rows compare MESI and \si protocols.
The table highlights the performance--com\-plex\-i\-ty tradeoff between MESI and \si,
and \si's performance advantage under false sharing.

This paper's goal is to get the advantages of performing self-invalidations and deferred flushes without most of the disadvantages and issues---achieving
(1) significantly lower complexity than MESI,
(2) performance and energy usage on par with MESI and significantly better than prior designs that use self-invalidations and deferred flushes,
and (3) out-of-the-box support for legacy programs.
To achieve this goal, we introduce
a novel, low-complexity approach to multicore cache
coherence called \emph{\barc}.
\Neat consists of two main design elements that contribute to its efficacy, one for each type of coherence action at synchronization operations.
First, \neat
uses novel lightweight mechanisms to reduce self-invalidation costs significantly by improving reuse of \emph{both} dirty and clean data across synchronization acquires,
which differs from prior work that relies on programmer annotations or inferred sharing patterns\cite{vips-directoryless-noc-coherence,denovo}).
Second, in \barc, writes to privately cached lines are not propagated until synchronization releases,
which differs from prior work that either uses MESI-style mechanisms to maintain ownership and propagate data at individual memory accesses~\cite{sarc-coherence}
or uses buffers to defer flushes (\ie, write-throughs or ownership registration requests) until overflowing buffers or reaching synchronization releases
~\cite{vips-directoryless-noc-coherence,denovo,denovond}.
\later{
\mike{Should the paper emphasize the second design element more? Maybe it's not such a big deal.
\rui{The second design element is two-sided: improving performance (due to better reuse of written data) by coalescing writebacks but causing bursty traffic at releases and having the overhead of per-byte write bits (similarly, DeNovo has fine-grained (per-word) registration bits).}}
}%


We perform two evaluations of \neat compared with state-of-the-art approaches.
First, we evaluate \neat's complexity
by implementing and verifying \neat and \mesi in the Murphi model checking tool~\cite{murphi}.
\later{
\mike{Is comparing with (some version of) DeNovo a possibility? Didn't we get the initial model from the DeNovo authors?
    \rui{Right we have the DeNovo model and should be able to use it for our evaluation. There are a couple of non-trivial revisions to the model though, to make it comparable with our \neat and \mesi models. Such revisions mainly include adding support for multiple words per line and using the same MSHR configuration and data race detection algorithm as the \neat model. \\
        But I was wondering if it'd be worth making the revisions and including DeNovo for only the complexity evaluation but not the performance evaluation. Wouldn't it make readers wonder why we compare with DeNovo only on complexity?}
    \mike{Good point.}}
}%
Our evaluation shows that \neat is about
an order of magnitude less complex to verify than the \mesi protocol.
Second, we implement a trace-based simulation of \barc,
compared with MESI and two self-invalidation-based coherence protocols called \emph{SARC} and \emph{\vips}~\cite{sarc-coherence, vips-directoryless-noc-coherence},
and evaluate on the PARSEC benchmarks~\cite{parsec-pact-2008}, three real server programs,
and the Phoenix benchmarks~\cite{phoenix}.
Our evaluation shows that \barc has competitive performance and energy usage
with MESI, and outperforms MESI significantly under false sharing.
\Barc also typically outperforms SARC and \vips.
\Barc reduces static power compared with MESI and SARC by eliminating the
coherence directory.
\later{
One of \barc's optimizations is also applicable to SARC, improving SARC to perform competitively with \barc;
however, SARC shares some of MESI's complexity including a directory and transient states.
}%
\Neat outperforms \vips because \neat's mechanisms are more effective in avoiding unnecessary self-invalidation than \vips's; 
\vips's optimizations are applicable to \neat, improving \neat's performance further.
\later{
\rui{Can we really claim that \sarc ``extends'' \mesi? I think it's just based on and simplifies \mesi to some sense.
\mike{Revised. Does SARC have transient states?}
\rui{Yeah I think so because of the Fwd-GetM message for SARC's writer coherence.}}
}%
These comparative results---summarized qualitatively in Table~\ref{tab:qualitative-comparison}---suggest
that \barc is a compelling alternative to MESI and state-of-the-art self-invalidation approaches
in terms of complexity, performance, and energy.
\later{\vignesh{Adding quantitative performance results -- improvements over VIPS and proximity to MESI -- here (and at the end of abstract) may help in highlighting the contributions. It
also presents a good opportunity to focus more on server-workloads (which benefit more)}}%

\section{Background: Self-Invalidation-Based Coherence Protocols}
\label{sec:background}


This section overviews state-of-the-art
cache coherence protocols that
self-invalidate readers' copies at acquire operations~\cite{sarc-coherence,vips-directoryless-noc-coherence, denovo, denovond, denovo-gpu, hlrc, quickrelease}.\footnote{Self-invalidation and deferred write-backs
originated as \emph{release consistency} mechanisms
for distributed shared memory
systems~\cite{lazy-release-consistency-dsm,comparison-entry-release-consistency,
midway-dsm, efficient-flexible-object-sharing, vmm-shrimp, accelerated-dsm}.}
\notes{
Depending on which cache lines are self-invalidated, some self-invalidations can be unnecessary and avoided.
}%
These protocols differ from each other mainly in their strategies for committing dirty lines to implement the data-value invariant:
\begin{itemize}[leftmargin=*]

\item \Sarc retains \mesi's directory to keep track of cache lines' ownership,
and a new writer initiates the write-back and invalidation of the old writer's copy~\cite{sarc-coherence} (Section~\ref{subsec:background-sarc}).

\item GPU coherence uses write-through caching for all data~\cite{denovo-gpu,quickrelease}, while \vips classifies private and shared data and uses write-back caching for private data and write-through caching for shared data~\cite{vips-directoryless-noc-coherence} (Section~\ref{subsec:background-vips}).
\later{
This paper's \neat protocol
(especially the baseline version as Section~\ref{subsec:unoptimized-protocol} describes) is, in spirit, like GPU coherence~\cite{denovo-gpu} (or \vips~\cite{vips-directoryless-noc-coherence}
without classification of private and shared data).
However, \neat makes important improvements over GPU coherence,
which lead to \neat's significantly better performance and energy efficiency compared with prior self-invalidation-based work.
}%
\item DeNovo and other DeNovo-based protocols rely on registering the ownership of dirty data in the LLC, rather than writing back dirty data directly~\cite{denovo, denovond, denovo-gpu, hlrc} (Section~\ref{subsec:background-registration}).

\end{itemize}
\notes{As discussed in Section~\ref{sec:intro}, assuming a DRF program,
either write-throughs (GPU coherence, \vips) or ownership registration (DeNovo and DeNovo-based protocols) can be buffered and delayed until buffer overflow or a release.}%

\subsection{Using MESI-Style Write-Backs}
\label{subsec:background-sarc}


\emph{SARC's} design retains part of MESI's directory to track ownership of dirty lines~\cite{sarc-coherence}.
A writer initiates the write-back and invalidation of the last writer's line.
SARC extends MESI by supporting \emph{tear-off} copies of lines for reads,
avoiding the need to maintain read-sharers in the directory.
A core's private cache self-invalidates a tear-off copy of a line at an acquire.
\Neat also avoids tracking sharers in the directory
and uses self-invalidation to ensure coherence.
Unlike SARC, \neat eliminates the directory and MESI protocol \emph{entirely},
and writes back all dirty bytes at releases to provide coherence.
\notes{
Empirically, the number of writebacks tends to be very low, avoiding a performance problem (Section~\ref{sec:eval}).
}%

\subsection{Using Delayed Write-Throughs}
\label{subsec:background-vips}

While CPUs use the complex MESI protocol or its variants,
GPUs prefer a simpler coherence protocol that the literature refers to as \emph{GPU coherence}~\cite{denovo-gpu,quickrelease}.
GPU coherence uses
self-invalidation at acquires and private write-through caches to keep the shared cache up to date.
To reduce the costs of write-through caching, the write-throughs can be buffered and delayed until the next release or buffer overflow.
This paper's baseline \neat design resembles GPU coherence in spirit, but the following important differences exist between the two designs:
\begin{itemize}[leftmargin=*]
    \item \Neat defers write-backs for \emph{all} dirty data, while GPU coherence uses buffers to hold outstanding write-throughs, which is subject to the capacity limits of the buffers.
\item \Neat introduces lightweight mechanisms that reduce unnecessary \si costs.
\end{itemize}
\later{\vignesh{Not sure if this point is worth mentioning above -- GPU coherence schemes also
have a different operating constraint than CPU coherence. GPU coherence schemes need to 
handle higher levels of outstanding memory requests (and can tolerate latency better). 
The implication of this design constraint is that any hardware buffering may need to 
support high bandwidth and may get filled up much more quickly than buffers in multi-cores.}}%

\noindent The above differences help \neat achieve significantly better performance and energy efficiency
than prior work.

Similar to GPU coherence,
\emph{VIPS}
uses
self-invalidation at acquires and delays write-throughs until a timeout,
a miss status handling register (MSHR) eviction, or a release operation~\cite{vips-directoryless-noc-coherence}.
\Neat avoids write-through costs imposed by VIPS by deferring all write-backs until releases.

To reduce performance costs due to unnecessary \sis,
VIPS optimizes \si by classifying pages as private or shared.
\later{Once a page is classified as some state, private or shared, each cached line of the page is marked as the same state as the page by setting a per-line state bit in the private caches.
Accesses to a private line do not trigger self-invalidation and do not write through.
When a private line becomes shared, its values are self-invalidated and written back.}%
For all shared pages, VIPS further distinguishes between read-only and read-write pages,
\later{and makes cache lines of read-only pages free from self-invalidation.}%
and self-invalidates only shared read-write pages.
While often beneficial, VIPS's classifications are sensitive to a program's data access patterns and have
limited impact on programs that mainly access data on shared read-write pages.
In contrast, our \Neat design introduces lightweight mechanisms to avoid unnecessary \sis without relying on specific data access patterns.
\later{regardless of how frequently the program accesses shared (read-only versus read-write) versus private pages.}%
\notes{
Since \neat's optimizations are applied at cache line granularity compared to page granularity,
\neat's optimizations are more effective to avoid unnecessary \sis.
\mike{That's not the only reason. That makes it sound simpler than it is. They're different optimizations.}
}%


\notes{
VIPS simulation. Rather than implementing MSHRs to enable non-blocking caches only to complicate the simulation, we simulate delayed write-through by buffering write-throughs until timeout, buffer being full (ie, “MSHR” eviction), or region boundaries. The buffer should be like a fully associative cache with some eviction policy (eg, LRU), with each of its entries storing the line address and per-byte dirty bits of write-throughs for a line. Since the buffer serves as MSHRs particularly for write-troughs, a reasonable size of such a buffer should somewhat smaller than normal MSHR sizes (say 8 assuming a normal MSHR size is 10 to 16), in order to reserve MSHRs for other types of outstanding mem requests.

A write-though leaves the buffer in the following 3 cases and triggers corresponding operations:
\begin{itemize}
    \item Timeout. Perform the actual write-through.
    \item Buffer being full. Evict a write-through as per the eviction policy.
    \item Encountering a region boundary. Stream all outstanding write-throughs in the buffer into the LLC with overhead charged due to streaming traffic and one-time LLC communication, as what we do with bulk write-backs for Neat.
\end{itemize}
Further, in the simulation, VIPS distinguishes between the two types of sync points --- lock releases and acquires --- like what Neat does to make a fair comparison (originally VIPS performs the same operations at whichever sync point).
}%

\subsection{Using Ownership Registration}
\label{subsec:background-registration}

DeNovo and DeNovo-based protocols rely on registering the ownership of dirty data~\cite{denovo,denovond,denovo-gpu,hlrc},
while \neat writes back dirty lines in bulk at releases.

\emph{DeNovo}
uses self-invalidation for out-of-date reads and \emph{registration} in the LLC to track a line's writers~\cite{denovo}.
Registration requires inclusion at the LLC and an extra level of indirection for writing back the data, hurting LLC capacity and introducing latency.
DeNovo further requires the program to be written in a deterministic, data-race-free language~\cite{dpj}.
\emph{DeNovoND} extends DeNovo to allow parts of a program to be nondeterministic~\cite{denovond},
but relies on the compiler to identify atomic accesses to
maintain coherence.
DeNovo and DeNovoND are thus not applicable to programs written in standard languages.

While much of the DeNovo line of work relies on programmer annotations
(or has other major differences from \neat, \eg, DeNovoSync~\cite{denovosync}; see Section~\ref{Sec:related}),
DeNovo and its ideas have been applied to GPUs
without requiring programmer annotations~\cite{denovo-gpu}.
This protocol uses registrations on dirty data; the registrations can be buffered and committed upon buffer overflow or at releases.
The protocol uses self-invalidations on all non-registered data at acquires to improve reuse of dirty data.

In contrast to the above DeNovo line of work, a \neat core writes back all dirty data to the LLC in bulk at releases, avoiding shared ownership metadata entirely.
\Neat uses lightweight mechanisms to improve reuse of both dirty and clean data without requiring programmer annotations.

\emph{Heterogenous lazy release consistency (hLRC}) builds on DeNovo, but exploits synchronization locality by registering only synchronization variables and lazily performing coherence actions
\later{(\ie, write-backs of dirty data and self-invalidations of valid data)}%
only when a remote synchronization operation is detected~\cite{hlrc}.
\later{hLRC thus exploits synchronization locality by performing write-back and self-invalidation only when synchronization variable ownership changes cores.
Compared with DeNovo, hLRC avoids expensive coherence actions at local synchronization, but fails to reuse written data across global synchronization operations.}%
\Neat does not detect synchronization locality since it targets CPU coherence where synchronization is mostly global, but it would be straightforward to extend \neat to exploit synchronization locality by avoiding coherence actions at detected local synchronization operations.

\medskip
\noindent
In summary,
researchers have explored various \si approaches to provide simple, efficient cache coherence
that avoids the complex directory and transient states of MESI~\cite{vips-directoryless-noc-coherence,denovo,denovond,denovo-gpu,hlrc}.
\later{~\cite{sarc-coherence,vips-directoryless-noc-coherence,denovo,denovond,denovosync,
    denovo-gpu,hlrc,spandex,quickrelease}.}%
Existing solutions for avoiding unnecessary \sis either depend on program data access patterns or require programmer annotations.
Further, in order to implement the data-value invariant,
these approaches either retain part of the directory to maintain shared ownership metadata, or incur write-through costs.


\newcommand*\BitNeg{\ensuremath{\mathord{\sim}}}
\newcommand*\BitAnd{\mathrel{\&}}
\newcommand{\cnt}{\code{CNT}\xspace}
\newcommand{\wbr}{\code{wbReceived}\xspace}

\section{The \Barc Coherence Protocol}
\label{sec:protocol}

This section first overviews \neat's design.
It then presents a \emph{baseline} version of \barc, which self-invalidates all private cache lines at acquires, and commits all dirty data at releases.
Finally, we present the full version of \barc that improves data reuse across synchronization operations, improving performance and energy
with acceptable additional complexity.


\subsection{\Barc Overview}
\label{subsec:protocol-overview}

\Barc is a set of modifications to a multicore processor that lacks support for cache coherence and core-to-core communication.
\later{This base multicore has
  no MESI protocol implementation, no coherence directory,
no coherence protocol and no support for core-to-core communication.}%
A core's cache associates with each cache line
a valid bit that is either valid (V) or invalid (I).
\later{
and a dirty bit that indicates updated line data.
\rui{Why do we assume dirty bits here? I'm concerned that readers may confuse it with the per-byte write bits.}
}%
Without loss of generality,
this section assumes that each core has a single-level private cache
and that all cores share a last-level cache (LLC).
It is straightforward to extend \barc's support to
multiple private cache levels (L1 and L2 caches), as in our performance and energy evaluation (Section~\ref{sec:eval-perf}).
Unlike common implementations of MESI or prior work that tracks ownership in the LLC~\cite{coherence-primer,sarc-coherence,denovo,denovond,denovo-gpu,hlrc},
\barc does \emph{not} require that the LLC be inclusive of the L1 caches.
\Barc does not require that the processor's interconnect supports core-to-core messages as implementations of MESI or prior work such as SARC and DeNovo do~\cite{coherence-primer,sarc-coherence,denovo,denovond,denovo-gpu,hlrc}.

\Barc does not explicitly maintain the \later{\emph{single writer, multiple readers} (SWMR)}%
SWMR invariant
maintained by MESI
and other protocols that use writer-initiated invalidation.
Instead, in \neat a core may read from out-of-date copies of lines,
and it may write to a line without immediately updating or invalidating other cores' valid copies of the line.
\Barc consequently provides coherence only for {\em data-race-free} (DRF) programs,
exploiting the DRF assumption that languages such as C++ provide~\cite{memory-models-cacm-2010,c++-memory-model-2008}.
Our work assumes that the compiler distinguishes synchronization operations
from regular memory operations in the compiled code
so that \barc operates with respect to synchronization from the original program source.

\Barc maintains the data-value invariant and ensures that, for DRF programs,
each read sees the value written by the last ordered (well-synchronized) write to the memory location.
To provide this guarantee, private caches self-invalidate
valid lines
and \cmt dirty data
at acquires and releases, respectively.
\later{
\begin{itemize}[leftmargin=*]

\item
Private caches \emph{self-invalidate}
valid lines at synchronization \emph{acquire} operations, e.g., lock acquire,
monitor wait, thread join, and a read of a C++ \code{atomic} or Java \code{volatile} variable.
Self-invalidation handles the fact that at an acquire operation, all lines are potentially out of date
and may need to be updated for a correctly synchronized program.
For example, consider the sequence of operations such as \code{$c_1$: wr x; $c_1$: rel; $c_2$: \textbf{acq}; $c_2$: rd x},
where $c_1$ and $c_2$ are cores performing read, write, acquire, and release operations.
The \code{\textbf{acq}} must invalidate $c_2$'s cached copy of \code{x} to ensure that the core reads the up-to-date value of \code{x} written by $c_1$.

\item
Private caches \emph{commit} dirty lines by writing back their written-to data at synchronization \emph{release} operations such as
lock release, monitor signal, thread fork, and a write to a C++ \code{atomic} or Java \code{volatile} variable.
Dirty lines need be written back at a release operation
so that other cores that perform well-synchronized reads of the line
can see up-to-date values.
For example, consider the sequence of operations
\code{$c_1$: wr x; $c_1$: \textbf{rel}; $c_2$: acq; $c_2$: rd x}.
The \code{\textbf{rel}} must make $c_1$'s update to \code{x} visible to $c_2$.
\end{itemize}
\smallskip
}%
Unlike prior work that buffers and delays write-throughs or registrations of dirty data until the buffer is full or a release is reached~\cite{vips-directoryless-noc-coherence,denovo,denovond,denovo-gpu,hlrc},
\neat does not rely on buffers and instead holds all dirty data in private caches until committing them at the next release.
A core's private cache keeps track of whether \emph{each byte} is dirty,
using one \emph{write bit} per byte. This feature is necessary for correctness:
non-dirty bytes may be out of date and should not in general be written back to the LLC.
This feature also optimizes \wbs by writing back only a line's dirty bytes. Note that
multiple (write) bits per cache line have been explored in prior work on precise
conflict detection~\cite{conflict-exceptions,arc} and sector caches~\cite{sector-cache-design}.

A core normally executes in a \emph{normal execution (NE)} state.
When a core performs an acquire operation, it transitions to the
\emph{self-invalidation (SI)} state, during which a core invalidates all of the valid lines in its private cache.
If any valid line is dirty, then the private cache writes back its dirty bytes.
Similarly, when a core performs a release, it
enters the \emph{commit (CM)} state, during which a core writes back all dirty bytes in its dirty lines to the LLC.

\begin{figure}[t]
  \centering
  \subfloat[The \mesi architecture with an inclusive directory.
Components added by \mesi are shaded grey.]{
\includegraphics[height=0.2\textheight]{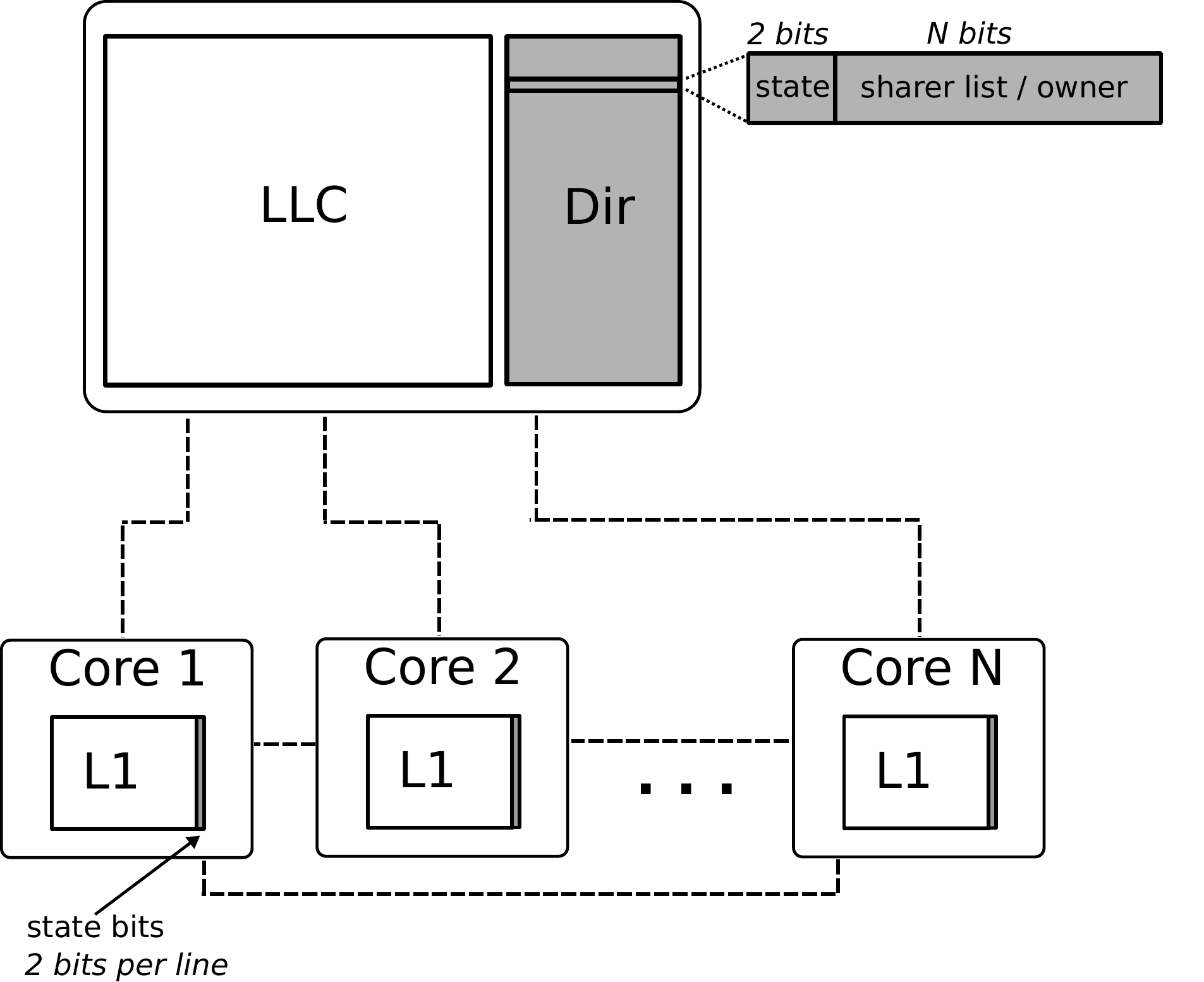} \label{fig:arch-mesi}}
  \medskip\\
\subfloat[The \neat architecture. Components added by \neat are shaded grey. \emph{Write sigs} = write signatures.]{
  \makebox[0.7\linewidth][c]{\includegraphics[height=0.2\textheight]{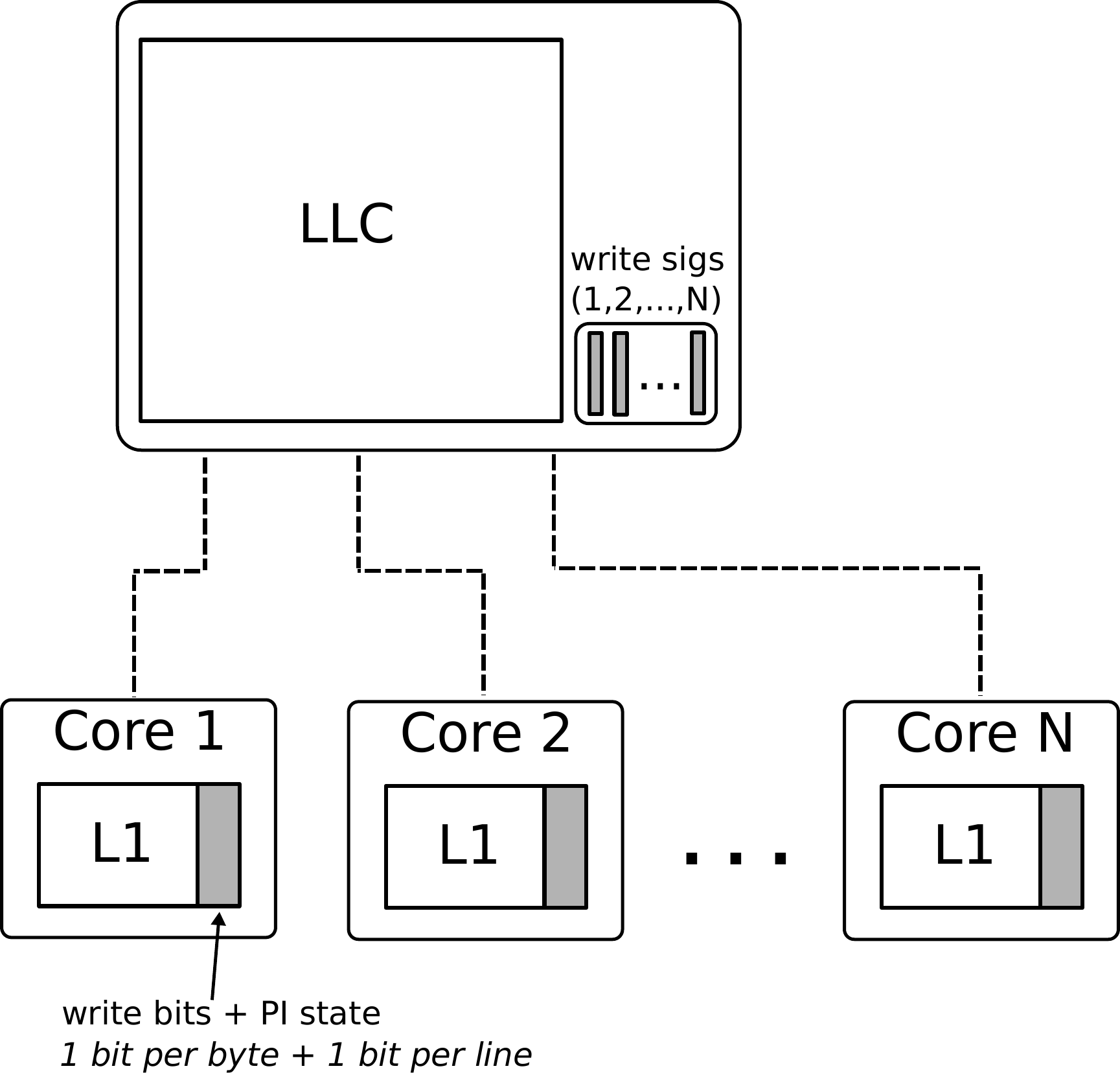} \label{fig:arch-neat}}}

  \caption{Illustrations of the \mesi and \neat architectures.
  \textnormal{Some components common to both designs, such as per-line valid bits, are omitted for simplicity.}}
  \label{fig:arch-cmp}
\end{figure}

Figure~\ref{fig:arch-cmp} shows the \mesi and \neat architectures that are both implemented on the base multicore processor,
where \mesi- and \neat-specific components are shaded grey.
Note that Figure~\ref{fig:arch-neat} includes support for mechanisms that avoid unnecessary self-invalidations and improve data reuse across synchronization operations---\emph{\wss} and \emph{PI state}---introduced and explained in Section~\ref{subsec:opts}.
Compared with \mesi, \neat eliminates the complex coherence directory and support for core-to-core communication,
and introduces relatively small structures (\wss) in the shared LLC and extra metadata for private cache lines.
\later{
Sections~\ref{sec:protocol-verification} and \ref{sec:eval-perf} evaluate the complexity,
performance, and energy of \neat, compared with \mesi. }%

\subsection{Baseline \Barc Protocol}
\label{subsec:unoptimized-protocol}

\begin{figure*}[t!]

\newcommand{\nan}{\multirow{2}{*}{N/A}}
\newcommand{\rbf}[1]{\multirow{2}{*}{\bf #1}}
\renewcommand\sfsmaller{\smaller}

\centering
\footnotesize
\subfloat[][\textbf{L1 controller's per-line state transitions.}]{
\begin{tabular}{@{}|c|c|c|c|c|c|c|@{}}
    \hline
    & \bf Read & \bf Write & \bf Replacement & \bf \Si & \bf \Cmt & \bf Data\\\hline
    \rbf I & read miss& write miss& \nan  & \nan & \nan & if write: set corresp.\\
    &  to LLC/- & to LLC/- &&&&  Wbs/V; if read: -/V \\\hline
    \multirow{3}{*}{\bf V} & \multirow{3}{*}{-/-} & set & if dirty: \wb (\cnt  = 1) & if dirty: \wb (\cnt  = 0) & if dirty: \wb (\cnt  = 0)& \multirow{3}{*}{N/A}\\
    & & corresp. & to LLC and clear all & to LLC and clear all &  to LLC and clear all & \\
    && Wbs/- & Wbs/I; if clean: -/I & Wbs/I; if clean: -/I & Wbs/-; if clean: -/- & \\\hline
\end{tabular}
\label{fig:unoptimized-protocol-l1}
}
\newline
\subfloat[][\bf L1 controller's core-wide state transitions.]{
\begin{tabular}{@{}|c|c|@{}}
    \hline
    \rbf{NE} & in addition to per-line actions and state transitions specified above,    \\
        & a) at a synchronization \emph{release}: -/CM; b) at a synchronization \emph{acquire}: -/SI \\\hline     
    \rbf{SI} & perform \textbf{Self-inv} on each valid line, send a data-less \wb message with\\ 
        & \emph{\cnt = total number of \wbs}, and wait for PutAllAck and all PutAcks/NE\\\hline
    \rbf{CM} & perform \textbf{Commit} on each dirty line, send a data-less \wb message with\\ 
        & \emph{\cnt = total number of \wbs}, and wait for PutAllAck/NE\\\hline
\end{tabular}
\label{fig:unoptimized-protocol-core}
}
\newline
\subfloat[][\textbf{LLC controller's state transitions.} \emph{Req} = the requesting core.
The LLC receives GetLine for each access miss
and data+Wbs+\cnt for each \wb from an L1.
\emph{\wbr} is a per-core counter at the LLC side that counts the \wbs by the LLC during a core's \si or \cmt.]{
\begin{tabular}{@{}|c|c|@{}}
    \hline
    \bf GetLine & \bf \Wb (data+Wbs+\cnt ) from core \\\hline
    \multirow{3}{*}{send Data to Req} & update line data corresponding to Wbs and \\
    & if \cnt = 0: \wbr{}++; if \cnt = 1: send PutAck to Req;\\
    & if \cnt $>$ 1: send PutAllAck to Req and reset \wbr to 0 when \wbr = \cnt\\\hline
\end{tabular}
\label{fig:unoptimized-protocol-LLC}
}
\later{
\mike{In (a), the V row (and also the PI row below) could be expanded to two rows, one for V and clean and another for PI and clean.
That'd avoid the conditionals inside of state transition cells. But that may be too big of a change for this submission.
\rui{Yeah it's not quite worth the time to make the change now.}}
}
\caption{Baseline \barc protocol.
  In each table, (non-bolded) entries show actions and state transitions, indicated by \emph{action/newState}
  with `-' indicating no action or changed state,
  in response to messages from other components (shown in \textbf{bold} column headers).
  The L1 controller's per-line and core-wide transitions take different actions and transitions
  depending on the current state (shown in bold row headers). Note that
  the protocol is described for a two-level cache hierarchy: core-private L1 cache and shared LLC.
  \emph{Wbs} = write bits.}
  \label{fig:unoptimized-protocol}
\bigskip\bigskip
\newcommand\new[1]{\textcolor{darkgreen}{#1}}
\centering
\footnotesize
\subfloat[][\textbf{L1 controller's per-line state transitions}.]{
\begin{tabular}{@{}|c|c|c|c|c|c|c|@{}}
    \hline
    & \bf Read & \bf Write & \bf Replacement & \bf \Si & \bf \Cmt & \bf Data\\\hline
    \rbf I & read miss & write miss& \nan &\nan & \nan & if write: set corresp.\\
    &  to LLC/- & to LLC/- &&&&  Wbs/V; if read: -/V \\\hline
    \rbf V & \multirow{2}{*}{-/-} & & if dirty: \wb  & \new{if contained in write} & if dirty: \wb & \nan\\
    & & set corresp.& (\cnt = 1) to LLC & \new{sig.: -/PI; otherwise: -/-} & (\cnt = 0) to LLC & \\\cline{1-2} \cline{5-5} \cline{7-7}
    \new{\rbf{PI}} &  \new{if read clean bytes: read miss} & Wbs/-& and clear  all Wbs/I; & \new{\multirow{2}{*}{-/-}} & and clear all Wbs/-; & \new{merge data at}\\
& \new{to LLC/-; if read dirty bytes: -/-} & & if clean: -/I & & if clean: -/- & \new{clean bytes/V} \\\hline
\end{tabular}
\label{fig:optimized-protocol-L1}
}
\newline
\subfloat[][\bf L1 controller's core-wide state transitions.]{
\begin{tabular}{@{}|c|c|@{}}
    \hline

    \rbf{NE} & in addition to per-line actions and state transitions specified above,    \\
        & a) at a synchronization \emph{release}: -/CM; b) at a synchronization \emph{acquire}: \new{send GetWrSig to LLC}/SI \\\hline     
    \rbf{SI} & \new{wait to receive write sig.,} perform \textbf{Self-inv} on each valid line, send a data-less \wb message with\\ 
        & \emph{\cnt = total number of \wbs}, and wait for PutAllAck and all PutAcks/NE\\\hline
    \rbf{CM} & perform \textbf{Commit} on each dirty line, send a data-less \wb message with\\ 
        & \emph{\cnt = total number of \wbs}, and wait for PutAllAck/NE\\\hline

\end{tabular}
\label{fig:optimized-protocol-core}
}
\newline
\subfloat[][\textbf{LLC controller's state transitions.} \emph{Req} = the requesting core.]{
\begin{tabular}{@{}|c|c|c|@{}}
    \hline
    \bf GetLine & \bf \Wb (data+Wbs+\cnt ) from core & \new{\bf GetWrSig} \\\hline
    \multirow{3}{*}{send Data to Req} & update line data corresponding to Wbs and & \new{send Req's write sig. to Req} \\
    & if \cnt = 0: \wbr{}++; if \cnt = 1: send PutAck to Req; & \new{and clear the write sig.} \\
    & if \cnt $>$ 1: send PutAllAck to Req and reset \wbr to 0 when \wbr = \cnt &\\\hline
\end{tabular}
\label{fig:optimized-protocol-LLC}
}


\caption{Full \barc protocol, with differences from Figure~\ref{fig:unoptimized-protocol} highlighted in \new{green}.
  \emph{Write sig.} = write signature.}
  \label{fig:optimized-protocol}
\end{figure*}



Figure~\ref{fig:unoptimized-protocol}
shows baseline \barc, which provides correct coherence for DRF programs and efficiently defers coalesced write-backs until synchronization operations.
Note that the basic protocol is inefficient because private caches invalidate all lines at acquires.
The table includes separate states and transitions for
(a) private cache lines,
(b) a core's private cache as a whole, and
(c) the LLC as a whole.
Note that \barc avoids the need for complex transient protocol states required by MESI,
and \barc adds no per-line state to the LLC (\eg, no directory; Figure~\ref{fig:arch-cmp}).


As in prior work (\eg,~\cite{denovo}), we assume each L1 controller has a
\emph{request buffer}, the storage array for which could be implemented as an explicit hardware buffer or using existing 
miss status handling registers (MSHRs). The request buffer tracks outstanding requests to the LLC.
The L1 controller adds a request buffer entry while waiting for a response from the LLC,
which is either a data request or a write-back message due to eviction of a dirty line,
and it removes the entry from the request buffer after receiving a response.
In \neat,
accesses to invalid (\emph{I}) and valid (\emph{V}) states in private caches are straightforward,
handled as misses and hits, respectively, except that a write sets corresponding write bits.
When a private cache evicts a \emph{clean} line
(\ie, a valid line without any write bits set),
it does so \emph{silently} (\ie, without communicating with the LLC).
When a private cache evicts a \emph{dirty} line (\ie, a valid line with at least one write bit set),
the private cache sends a write-back message
to the LLC that includes the dirty bytes and corresponding write bits, and a count value \emph{\cnt{}=1} (detailed below), which
directs the LLC to send back an \ack message \emph{PutAck} immediately after receiving this one write-back message.
The private cache changes the line state to I immediately, and
the core may continue execution in general (as long as there is no dependency with the pending requests in the request buffer),
but the core must wait for acknowledgments of all outstanding write-backs
before executing operations \emph{after} the next release.
An access miss or write-back message does not lead to transient states for cache lines because there are no conflicting access requests
forwarded from other cores.
\Neat does not risk deadlock either because the LLC responds to an access miss or write-back directly without relying on any responses from other cores.

\later{Besides the above actions during a core's normal execution in the \emph{NE} state,}%
A core transitions to a \cmt (\emph{CM}) state at a release and
to a \si (\emph{SI}) state at an acquire.
While in the CM or SI state, a core does not fetch or execute instructions,
and instead waits for the L1 controller to perform operations iteratively over all dirty lines (if in CM state) or valid lines (if in SI state).

During the CM state, the L1 controller writes back all its dirty lines and avoids per-line acknowledgment of each \wb by
having the LLC send a single \emph{PutAllAck} message once it has received all \wbs.
To facilitate the LLC acknowledging such bulk \wbs, each \wb message has an integer field \emph{\cnt} that indicates the number of \wbs
that the LLC should receive before sending back an acknowledgment.
The L1 controller counts the number of \wbs when committing dirty lines; after sending all \wbs,
the L1 sends an extra \wb message to the LLC with \cnt set to the number of write-backs and no data.
\later{
\mike{I thought we decided that the L1 could send the count with the last \wb message during CM?
I don't want to just change it here since it's not essential and it might require protocol table or other changes, too.
\rui{Yeah that's what we decided, but it's easier for us to use an extra data-less message in the protocol table.}}
}%
The LLC knows to wait before acknowledging \wbs because the L1 controller sets \cnt to 0 for \wbs sent during the CM state.
The LLC maintains per-core counters \emph{\wbr} that count the \wbs received from each core with \cnt{}=0.
\later{Upon receiving a \wb with its \cnt equal to 0, the LLC only increases \wbr for the core rather than sending any acknowledgment.}%
After receiving the extra \wb with a \cnt greater than 0, the LLC compares the \cnt and \wbr of the same core;
if the two are equal, it sends back a PutAllAck message to acknowledge all of the core's \wbs and clears the \wbr counter.
We support the more general case of an out-of-order network between cores and the LLC,
so a core's bulk \wbs may be reordered with its extra \wb, allowing the LLC to receive a \cnt greater than \wbs from the core so far.
In such a case, the LLC keeps waiting for more \wbs to arrive from the core, incrementing \wbr when appropriate.
To ensure that a core's dirty data are correctly written back and become visible at releases,
a core in the CM state does not transition back to the NE state until it receives acknowledgments (PutAllAck and any PutAcks) for all outstanding \wbs.

We assume a centralized LLC for the above discussion, but it is straightforward to apply \neat's bulk \wbs to a distributed LLC. 
With a distributed LLC, during the CM state, the L1 controller counts the number of \wbs it sends to each LLC bank and sends an extra \wb message to each bank with the corresponding count number.
Each bank acknowledges receiving all the \wbs by sending back a PutAllAck message.
A core in the CM state does not transition back to the NE state until it receives PutAllAcks from all LLC banks. 
To handle dropped packets on an unreliable network, if a core does not receive a response within a time limit, it should retry committing write-backs from the beginning.

During the SI state, the L1 controller invalidates all valid lines.
It writes back dirty data for any dirty lines that it invalidates,
in the same way as the CM state's \wbs, avoiding per-line \ack of each \wb.
After invalidating all valid lines, the L1 controller transitions back to the NE state only after receiving a PutAllAck message.
This behavior preserves that each core sees its own last write even in the context of an out-of-order network.
Note that the L1 controller does not need to wait for outstanding \emph{PutAck} responses (for \wbs due to \emph{dirty evictions})
before transitioning from SI to NE
because the corresponding write-backs occupy request buffer entries while waiting for acknowledgements.
\later{
\cnt's value applies generally to all \wb messages, including bulk \wbs performed at synchronization operations and individual \wbs due to dirty evictions.
On a dirty eviction, the L1 controller sets the \wb message's \cnt to 1 to inform the LLC that a \emph{PutAck} message should be sent back immediately
when the LLC receives the \wb.}%


\subsection{Full \Neat Protocol}
\label{subsec:opts}



Conservative self-invalidation is expensive in terms of run-time performance
\later{ traffic to LLC,}%
and energy because it
conservatively invalidates lines that may be out of date,
which can hurt cache locality and result in avoidable cache misses.
We introduce two mechanisms that significantly help reduce the costs of self-invalidation and improve data reuse across acquires.
Figure~\ref{fig:optimized-protocol} shows the corresponding full version of the \barc protocol.

\subsubsection{Partially invalid state}

We observe that, at an acquire operation, a core can \emph{delay} self-invalidating a line
until its next read access to any \emph{clean} byte in the line.
In other words, a core can write to the line or read a byte that the core has already written,
since these values will be up to date for a DRF execution.
To make use of this observation, we introduce a \emph{partially invalid (PI)} state for private cache lines to indicate that a line may have out-of-date data in its \emph{clean} bytes.
During \si,
instead of invalidating each valid line (\ie, changing its state to I),
the private cache \emph{partially invalidates} each valid line, by changing its state to PI.
Any subsequent write to a PI line is a hit, and
a read to a dirty byte of a PI line is also a hit.
But a read to clean byte(s) of a PI line is a miss;
the private cache fetches updated value(s) of the byte(s) from the LLC, merges them into the private cache line
(overwriting only clean bytes), and marks the line valid (\ie, PI $\rightarrow$ V).

Note that with this mechanism, while an L1 controller is in the SI state
(\ie, at an acquire)
it does \emph{not} clear any line's write bits, regardless of whether
the line is partially invalidated or left valid (a line may remain valid after self-invalidation only if using the write signature mechanism, described below).
In contrast, during the CM state (\ie, at a release),
the L1 controller clears a line's write bits when writing back dirty bytes,
and does \emph{not} invalidate or partially invalidate any lines.
In a DRF program, it is correct for a core to read from its own bytes at least until an acquire operation.

The PI state has some similarities to DeNovo's \emph{touched bit}~\cite{denovo}, but
there are some key differences.
The touched bit indicates that a word is exclusively read by the current core and
the core has up-to-date data for this word at the end of the current parallel phase.
The PI state indicates the current core may have stale data for those bytes that were not written by the core.
The touched bit is at word granularity while the PI state is at line granularity.
Finally, DeNovo's use of the touched bit relies on programmer annotations, while Neat's use of the PI state does not rely on annotations.

\subsubsection{Per-core write signatures}

Our second observation is that a core $c$ can \emph{skip} self-invalidating a line
if that line has not been updated in the LLC by another core since $c$'s last acquire operation.
To identify such lines, \barc maintains per-core \emph{write signatures}
at the LLC.
A core $c$'s write signature indicates which of $c$'s private lines were updated in the LLC \emph{by any other core}
since $c$'s last acquire operation.
Each write-back to an LLC line by core $d$ adds the line (address) to the write signatures for all cores \emph{other than} $d$.

A core fetches its write signature at the start of \si and only invalidates (or partially invalidates) those lines contained in the write signature.
The LLC clears a core's write signature once it services the core's fetch request for the \ws.
We note that a core fetches its write signature only after the core succeeds on an acquire, so under the DRF assumption, there is no race on write signatures. 

The write signature mechanism is important mainly when acquire operations and thus \sis are frequent.
When acquire operations are infrequent, \sis are less frequent, and their costs tend to amortize
better over other execution costs.
We can thus optimize for frequent self-invalidations by making \wss small,
which saves time and area by avoiding sending and storing large \wss.
Our implementation uses Bloom filters~\cite{bloom-filter} for over-approximated write signatures
and sends compressed versions of sparsely populated write signatures (Section~\ref{sec:eval-perf}).


\later{
\rui{"Applying optimizations to VIPS or MESI: is there a reason why the same optimizations cannot be applied to either VIPS, SARC, or MESI?  I realize and appreciate that applying them to MESI is harder, but is there a fundamental reason why they cannot be applied?  This would eat further into the already small gains of Neat.  Moreover, this does lead into the classic question for work in this area: given the very small gains in terms of performance and energy over MESI, why would a company want to adopt this solution?"\\
We had evaluated applying Neat's write signature optimization to SARC and observed perf improvement. Similarly, the optimization can be applied to VIPS. I don't understand how the optimizations can be applied to MESI, since MESI doesn't do self-invalidations. As for the adoption issue, we argue that Neat is simpler than MESI, but has comparable performance to MESI.}}%

\section{Evaluation of Correctness and Complexity}
\label{sec:protocol-verification}

We validated \barc's correctness and
quantitatively evaluated \neat's complexity, directly comparing to the complexity of a \mesi protocol implementation,
in the widely used \emph{Murphi} model checking tool~\cite{murphi}.
\later{
\rui{"why didn't you include DeNovo in your comparison (since the files theoretically would have been included in the same release as the MESI ones)?"\\
I don't remember why we didn't include DeNovo in the complexity evaluation, but we should be able to do that.
\mike{Is it because DeNovo requires annotations of which locations will be accessed in a region?
What would that mean for the comparison in Murphi?}
\rui{They only modeled one address and one region in Murphi, so they actually assumed no programmer annotation and just let Murphi explore various combinations of reads and writes in a region as long as DRF is guaranteed. I'll modify their specification to make a comparable version to our \neat/\mesi specifications.}}}%

\subsection{Methodology}

We implemented MESI in Murphi based on the GEMS two-level MESI protocol~\cite{gems}.
\notes{
We obtained a Murphi specification for \mesi used in prior work~\cite{denovo-verification}.\footnote{We
have not compared with the specification for DeNovo~\cite{denovo-verification} because (1) it supports
only one line and byte per line, and (2) DeNovo annotation's requirement makes it not directly comparable to \neat.}
\mike{Rewrote above. Please check.}
The provided \mesi specification is based on the GEMS two-level MESI protocol~\cite{gems} and models a single line with a single byte.  We extended 
the model to support a cache with multiple lines and bytes.
}%
We implemented \barc in Murphi by directly specifying the protocol described in Section~\ref{sec:protocol}.
For both \mesi and \barc, our specifications model two cores, a two-level cache hierarchy (\ie, private L1s and a shared LLC),
an unordered network with unlimited capacity,
two data values, and up to two lines and two bytes per line.
In addition to modeling standard cache operations due to reads, writes, and evictions of lines,
the \mesi and \barc specifications model acquire and release operations.

\later{
\mike{I think we need to say more about the specifications,
\eg, what the operations are: reads, writes, acq, rel operations.
\brandon{Agreed. More detail would help.  We should also say, wherever it fits, that we'll release our model.}
\mike{Added something about the first above, but it could be improved. Added the second part below.}}
}%
Verification using a Murphi model
serves two purposes: providing a demonstration that the protocols are correct, and allowing a comparison of the complexity of the protocols.
We will make our \barc model and modified \mesi model publicly available so that others can reproduce and modify our evaluation
of correctness and complexity.

To check correctness, both specifications check a ``last-write'' assertion for each state explored by Murphi:
any read by a core of a byte in a cache line should see the same value written by the last write by any core.
(The specifications also check protocol-specific assertions, \eg, the \mesi specification checks
the ``single writer'' invariant that there is at most one modifiable copy of each line at a time.)

We note that for \neat the last-write property holds only for
\emph{data-race-free} (DRF) executions; \neat has undefined behavior for data races.
\mesi by contrast provides the last-write property for all executions---but
real systems that use \mesi typically apply compiler and hardware optimizations
allowed by language and hardware memory models that yield undefined or weak semantics for executions with data races~\cite{memory-models-cacm-2010}.
It thus makes sense to verify \neat and \mesi for DRF executions only.
To limit verification to DRF executions only, we extend the \neat and \mesi specifications to detect a subset of data races
for which a byte's last write may be undefined, \ie, any access to a byte is not ordered by synchronization with the prior write of that byte.
That is, any read or write to a byte by core $c_2$ that is preceded by a write to the same byte by a different core $c_1$
must be well synchronized, \ie, there must exist a sequence of operations \code{$c_1$: wr x; $c_1$: rel; $c_2$: acq; $c_2$: rd/wr x},
where \code{rel} and \code{acq} are any release and acquire operations, respectively.
\later{
(Note that the specifications allow read--write races, which do not lead to undefined behavior for \barc.)
\brandon{The parenthetical here is little bit of a ``huh??''  Do we introduce this fact earlier in the paper?
If not, we should.  Once we do, we can do a back-reference here to make this easier to understand in passing here.}
}%
If the \mesi or \neat specification detects a violation of this property in an execution,
it transitions the execution state to the \emph{initial} state (\ie, the state before execution starts),
effectively cutting off exploration of racy states.

\begin{table*}[t]
\renewcommand{\code}{}
\newcommand{\lots}{Timeout}
\newcommand\heading[2]{\bf #2}
\newcommand\choice[2]{#2}
\centering
\begin{tabular}{@{}c@{$\;\;$}c@{$\;$}|r@{$\quad\;$}r@{$\quad\;$}r@{$\quad\;$}r@{}}
    \heading{@{}c}{Lines} & \heading{r|}{Bytes/line} & \heading{r}{\mesi} & \heading{r}{\bcunopt} & \heading{r}{\bcpi} & \heading{r@{}}{\bcbf{}}\\\hline
    \bf 1 & \bf 1 & 29,050 & 933 & 7,056 & 22,557\\
    \bf 1 & \bf 2 & 402,911 & 13,231 & 130,506 & 391,755 \\
    \bf 2 & \bf 1 & \choice{\lots}{739,314,210} & 92,736 & 3,497,049 & 35,952,474\\
    \bf 2 & \bf 2 & \lots & 36,260,993 & \lots & \lots
\end{tabular}
\vspace*{0.5em}
\caption{States explored by Murphi for \mesi and various \barc specifications,
for 1--2 lines and 1--2 bytes per line.
\textnormal{For all configurations, Murphi explores only data-race-free states.}}
\label{tab:states:drf-only}
\label{tab:states}
\end{table*}

\later{
\begin{table}[t]
\renewcommand{\code}{}
\newcommand{\lots}{Timeout}
\newcommand\heading[2]{\bf #2}
\newcommand\choice[2]{#2}
\centering
\subfloat[][Model checking explores data-race-free executions only.]{
\begin{tabular}{@{}c@{\quad}c|r@{$\quad\;$}r@{$\quad\;$}r@{$\quad\;$}r@{}}
    \heading{@{}c}{Lines} & \heading{r|}{Bytes/line} & \heading{r}{\mesi} & \heading{r}{\bcunopt} & \heading{r}{\bcpi} & \heading{r@{}}{\bcbf{}}\\\hline
    \bf 1 & \bf 1 & 29,050 & 933 & 7,056 & 22,557\\
    \bf 1 & \bf 2 & 402,911 & 13,231 & 130,506 & 391,755 \\
    \bf 2 & \bf 1 & \choice{\lots}{739,314,210} & 92,736 & 3,497,049 & 35,952,474\\
    \bf 2 & \bf 2 & \lots & 36,260,993 & \lots & \lots
\end{tabular}
\label{tab:states:drf-only}
}

\later{
\mike{The jump in \# of states from \bcunopt to \bcpi is enormous---just from adding the PI state (right?). Does that make sense?
Update: We discussed it and it makes somewhat more sense, although I think it's still worth looking into.}
}

\subfloat[][Model checking explores all executions.]{
\begin{tabular}{@{}c@{\quad}c|r@{$\quad\;$}r@{$\quad\;$}r@{$\quad\;$}r@{}}
    \heading{@{}c}{Lines} & \heading{r|}{Bytes/line} & \heading{r}{\mesi} & \heading{r}{\bcunopt} & \heading{r}{\bcpi} & \heading{r@{}}{\bcbf{}}\\\hline
    \bf 1 & \bf 1 & 12,289 & 1,371 & 11,175 & 37,233\\
    \bf 1 & \bf 2 & 79,609 & 13,911 & 172,482 & 496,342 \\
    \bf 2 & \bf 1 & \choice{\lots}{117,795,838} & 80,199 & 3,101,617 & 37,324,500\\
    \bf 2 & \bf 2 & \lots & 19,338,610 & \lots & \lots
\end{tabular}
\label{tab:states:all}
}
\vspace*{0.5em}
\caption{The number of states explored by Murphi for \mesi and various \barc specifications,
for 1--2 lines and 1--2 bytes per line.
\textnormal{For the specifications evaluated in (a), Murphi explores only data-race-free states for which \barc provides guarantees.
For the specifications evaluated in (b), which we include for completeness, Murphi explores states with data races for which \barc does not provide guarantees. Fewer states indicate a lower protocol complexity.}}
\label{tab:states}
\end{table}
}

\subsection{Protocol Complexity Results}
\label{subsec:complexity-results}

Table~\ref{tab:states:drf-only} reports how many execution states Murphi explored
for the \mesi specification and three variants of the \neat specification.
\emph{\bcunopt} is the baseline \neat;
\emph{\bcpi} includes the partially invalid (PI) state mechanism; and
\emph{\bcopt} is the full version of \neat and includes the PI state and write signature mechanisms.
For each configuration, the table reports states explored by Murphi
for the four combinations of 1--2 lines and 1--2 bytes per line.

As the results show, \mesi's complexity (i.e., number of states) increases more quickly than \barc's complexity as the number of lines in the cache increases.  This is because \mesi requires per-line transient states to keep the lines coherent.
In contrast, \barc's complexity increases more quickly than \mesi's as the number of bytes per line increases, because \barc maintains per-byte write bits for private cache lines.
However, \barc's relative complexity for an additional byte is lower
than \mesi's relative complexity for an additional line,
since \mesi's per-line state space is considerably larger than \barc's per-byte state space: \mesi adds many transient states per line, while \barc adds a single write bit per byte.
\notes{\vignesh{Does the last sentence mean that we have one transient state per byte in \barc? Stating the precise number of
transient states in \mesi and \neat may be helpful here.
\mike{In what way is it a \textbf{transient} state?
I guess it's true that one could consider there to be 2$^{64}$ possible states for a cache line's write bits,
but I don't think we want to present it that way.}
\vignesh{Actually, I am not sure what the number of transient states should be for \barc. My confusion
was more because the last sentence seemed to compare objects of different types -- transient states in \mesi
with bits in \barc.}
\mike{I believe ``transient states'' are states needed because of protocol races, and so \Neat doesn't have transient states.}}}%

The verification state space is large for a \barc or \mesi configuration with two lines and two bytes per line.
Except for \bcunopt, Murphi never completed checking any of the configurations (indicated by \emph{Timeout} in the table),
even after exploring hundreds of millions of states, which took several days running on a
\later{
\brandon{which tier?
\mike{Couldn't figure this out easily. But in general we should switch to Standard Tier to save some \$.}}
}%
VM hosted by Google Compute Engine.
\later{
\vignesh{If it easy to find, we should report main memory available on the VM. COUP~\cite{coup-micro-2015} 
did the same while reporting unfinished MURPHI runs
\mike{Rui doesn't know since it was a long time ago.}}
}%



\later{
Considering that a normal cache has significantly more lines than bytes within a line,
we expect \barc to have more significant advantage over \mesi in real world than the simplified verifications.
\mike{How would this have a real-world advantage? I don't think there's a ``real-world verification'' that would try to model a real-size cache. It wouldn't scale.}
}%


The main takeaway from these results is that \barc is considerably less complex than \mesi. \Mesi and the full \barc configurations
with just one line have comparable numbers of states, with \barc having somewhat fewer states.
For the two-line, one-byte configuration, \neat has \textbf{20\boldmath$\times$} fewer states than \mesi,
due to \barc's small per-line state space.



\later{
For completeness, Table~\ref{tab:states:all} shows states explored by Murphi for the same configurations of \mesi and \neat---but
modified to explore all execution states including those with data races.
\Neat would fail last-write assertions in this context,
so the configurations disable last-write tracking and assertion checking for both \mesi and \neat.
These configurations also do not track state to detect last-write races.
Exploring execution states with races increases the state space,
while eliding last-write tracking and race detection decreases the state space
(Murphi inherently treats all specification states as part of the state space that it explores),
leading to a mix of increased and decreased states compared with Table~\ref{tab:states:drf-only}.
We provide these additional results to further characterize \barc's verification complexity.
Although the data show an increase in states for \barc,
the experiment explores states for which \barc has undefined behavior,
so the results are \emph{not} indicative of \barc having higher complexity than \mesi.
\vignesh{I was wondering if all-executions results merit a table and a paragraph. Not sure what the significance of 
reachable states is if we disable last-write tracking and assertion checking in both \mesi and \neat. IOW, what property
are we verifying for racy executions? (I can see SWMR for \mesi but not clear what we are checking for \neat).
\mike{Agreed. Removed it.}}
}%


\later{
\mike{The following seems redundant, particularly with the ``all executions'' paragraph removed.}
In summary, in addition to verifying that the \neat protocol is correct with respect to key invariants,
these results show that \neat is less complex than \mesi because \barc's verification state space has fewer states than \mesi's space.
}%

\section{Evaluation of Performance and Energy}
\label{sec:eval-perf}

This section evaluates the performance and energy usage of \neat,
compared with
a state-of-the-art \mesi protocol implementation~\cite{illinois-mesi,coherence-primer} and
two self-invalidation-based protocols from the literature~\cite{sarc-coherence, vips-directoryless-noc-coherence}.

\subsection{Implementation and Methodology}
\label{sec:eval:perf:impl-meth}

Our experiments measure run-time performance and energy
consumption,
\later{and interconnect bandwidth consumption}%
using the RADISH simulator~\cite{radish} modified to
implement (1) \neat, (2) a directory-based MESI protocol
implementation~\cite{coherence-primer}, and two self-invalidation-based protocols,
(3) \emph{SARC}~\cite{sarc-coherence} and (4) \emph{VIPS}~\cite{vips-directoryless-noc-coherence}.
We will make our simulation and modeling infrastructure publicly available.
All simulator backends consume the same trace of instructions from a PIN-based front end~\cite{pin}. 
Each core has a two-level private cache hierarchy, and the LLC is backed by off-chip main memory.
A core's L2 is inclusive of its L1.
The LLC is not inclusive of private caches for \sarc, \vips, and \neat (and need not be),
but the LLC is inclusive of the L2 for \mesi to support an inclusive directory cache embedded in the LLC~\cite{coherence-primer}.

\begin{table}[t]
  \renewcommand\sfsmaller{}
  \centering
  \begin{tabular}{@{}lp{6cm}@{}}
    \bf \multirow{2}{*}{Processor} & 32-core chip at 1.6 GHz. Each non-memory-access instruction takes 1 cycle. \\\midrule
    \bf \multirow{2}{*}{L1 cache} & 8-way 32 KB per-core private cache, \newline 64 B line size, 4-cycle hit latency \\\midrule
    \bf \multirow{2}{*}{L2 cache} & 8-way 256 KB per-core private cache, \newline 64 B line size, 10-cycle hit latency \\\midrule
    \bf Remote core & \multirow{2}{*}{15-cycle one-way cost ~~(\mesi only)} \\
    \bf cache access \\\midrule
    \bf \multirow{2}{*}{LLC} & 64 B line size, 50-cycle hit latency, \\
            & 32-way 64 MB shared cache \\\midrule
    \bf On-chip & \multirow{2}{*}{16-byte flits, 100 GB/s bandwidth} \\
    \bf interconnect \\\midrule
    \bf Memory & 120-cycle latency \\
  \end{tabular}
\vspace*{0.5em}
  
  
  \caption{Architectural parameters used for simulation.}
  \label{tab:arch-params}
\end{table}


Our simulators model and measure execution cycles and on-chip traffic,
using parameters shown in Table~\ref{tab:arch-params}.
The simulators model single-issue, in-order cores in which non-memory instructions have an IPC of one,
and an interconnect network that uses 16-byte flits.
The \neat and \vips simulators model the cycle cost of performing \si and \cmt (either bulk write-backs by \neat or delayed write-throughs by \vips) at synchronization operations
based on the total size of messages sent and the bandwidth available between a core and the LLC.
We report the maximum cycles of any core as execution time.

To measure energy consumption, we use the \emph{McPAT} energy modeling tool~\cite{mcpat}, providing it with the output statistics from our simulator.
We report total energy for the \emph{cache and memory subsystem}, including the on-chip interconnect and LLC-to-memory communication.
We exclude reporting energy for operations within the cores because our simulator does not
collect detailed core-level statistics such as ALU and branch instructions, which McPAT needs to compute a core's energy usage.
However, these excluded operations and thus the excluded energy should be identical across all configurations.

We use McPAT to model energy of all \mesi, \neat, \sarc, and \vips components
across the cache and memory subsystem, with one exception:
It is unclear how to model \neat's Bloom-filter-based \wss (Section~\ref{subsec:opts}) in McPAT,
\later{When servicing a core's fetch request for the Bloom filter,
    the LLC sends bit positions instead of a bit vector if it is more compact.}%
so we estimate the per-access energy of Bloom filters using values reported by prior work~\cite{l-cbf}.
Each core's \ws is a 1008-bit Bloom filter (which fits in eight 16-byte flits including a control message).
Specifically, we assume L-CBFs, each with 1008 1-bit counters, and derive their per-operation dynamic energy by assuming linear relationships
between the per-operation energy and entry count (as well as count width) for an L-CBF.
We compute total dynamic energy due to \ws operations by multiplying the per-operation energy by the numbers of \ws operations counted in the simulator.

\begin{table}[t]
    \centering
      \renewcommand\sfsmaller{\small}
      \newcommand{\na}{N/A}
      \newcommand{\cmk}{\ding{51}}
  \newcommand{\xmk}{\ding{55}}
        \begin{tabular}{@{}l|l|l@{\qquad}l@{\qquad}l@{}}
          &                        & \multicolumn{3}{c@{}}{\bf Mechanisms} \\
          \textbf{Category} & \textbf{Config} & \textbf{PI} & \textbf{WS} & \textbf{CLA} \\\hline
          Non-self-inv. & \mesicfg~\cite{illinois-mesi} & \na & \na & \na \\\hline
          \multirow{3}{*}{Prior self-inv.}& \scunopt~\cite{sarc-coherence} &  \xmk & \xmk & \xmk \\
          & \vipsunopt &  \xmk & \xmk & \xmk \\
          & \vipspsrw~\cite{vips-directoryless-noc-coherence} &  \xmk & \xmk & \cmk \\\hline
          \multirow{4}{*}{\Neat} & \ntunopt &  \xmk & \xmk & \xmk \\
          & \ntpi & \cmk & \xmk & \xmk \\
          & \ntopt & \cmk & \cmk & \xmk \\
          & \ntpsrw & \cmk & \cmk & \cmk \\
          
        \end{tabular}%
\vspace*{0.5em}
    

      \caption{Simulated configurations and the mechanisms they employ to avoid unnecessary \si.
      \textnormal{\emph{PI} = \neat's partially invalid state.
      \emph{WS} = \neat's \wss.
      \emph{CLA} = \vips's page classifications.}}
      \label{tab:configs}
    \end{table}
    

\subsubsection*{Evaluated configurations}

Table~\ref{tab:configs} shows the configurations we evaluate.

\emph{\ntunopt} is baseline \neat;
\emph{\ntpi} includes the partially invalid (PI) state mechanism; and
\emph{\ntopt} is the full version of \neat and includes the PI state and write signature mechanisms.
\emph{\ntpsrw} is \ntopt plus two page-level classification optimizations used by VIPS:
private vs.\ shared pages, and read-only vs.\ read-write pages~\cite{vips-directoryless-noc-coherence} (Section~\ref{subsec:background-vips}).

We evaluate a configuration called \emph{\scunopt} based on the design described by prior work~\cite{sarc-coherence}.
\later{Like the SARC paper's evaluation~\cite{sarc-coherence},
our SARC implementation treats all read-only lines as tear-off copies.
As a result, the directory does not include any sharer information.}%
We also implemented and evaluated an idealized (perfect) implementation of SARC's writer prediction~\cite{sarc-coherence},
but found that it had negligible performance impact, so we exclude writer prediction from the evaluation for simplicity.

\notes{
\mike{We've removed \sarcws from the results, so excluded paragraph about it below.
We could potentially still mention that we tried applying the \ws optimization to \sarc,
but then it'd be kinda weird that we don't show results.}
To demonstrate the generality of \barc's \ws optimization,
we also evaluate a \cfg that is a hybrid of \sarc and \neat, called \emph{\sarcws}.
This configuration applies \barc's \wss to \scunopt:
\sarcws uses \wss to avoid self-invalidating privately cached lines unmodified in the LLC since the core's last self-invalidation.
While the results show that \sarcws often performs comparably with \bcopt,
\mike{Still true? In any case, could omit the first part of the sentence.
\rui{Not for servers. How about we commit the \sarcws results to emphasize the \ntopt and \ntcla results.}
\mike{Huh? Do you mean \textbf{omit} instead of commit?}}
\sarcws (and \scunopt) are arguably more complex than \bcopt,
since SARC builds on top of a MESI baseline (albeit a MESI baseline that does not track sharers)~\cite{sarc-coherence}.
}%

In contrast with \neat and \sarc, \vips uses write-through caches;
MSHRs buffer write-throughs until they run out, or a timeout or synchronization release is reached~\cite{vips-directoryless-noc-coherence}.
To approximate the MSHRs' behavior, our simulator implements a \emph{write-through buffer}
with an LRU eviction policy and an infinite timeout.
\later{Consequently, the \vips simulator writes back to the LLC only at a release or write-through buffer eviction.}%


\Vips classifies private and shared memory pages and read-only and read-write pages to avoid unnecessary \si~\cite{vips-directoryless-noc-coherence} (Section~\ref{subsec:background-vips}).
To understand the effectiveness of \vips's classifications compared to \neat's self-invalidation-reducing mechanisms,
we evaluate two \vips \cfgs, \emph{\vipsunopt} and \emph{\vipspsrw}.
\vipsunopt does \emph{not} classify private and shared pages or read-only and read-write pages,
but writes through and self-invalidates all lines, effectively treating all lines as shared and read-write.
\vipsunopt is like \ntunopt in how both protocols commit and self invalidate cache lines at synchronization operations,
though \vipsunopt is subject to the limited capacity of the write-through buffer (10 entries) and
does not wait until the next release to commit \emph{all} dirty lines as the \neat configurations do.
We note that \vipsunopt represents so-called ``GPU coherence'' (Section~\ref{subsec:background-vips}).
\emph{\vipspsrw} \later{, which represents optimized \vips~\cite{vips-directoryless-noc-coherence},}%
adds the classifications to \vipsunopt,
writing through and self-invalidating only those lines that are marked as shared and read-write.
For lines that are marked as private, \vipspsrw writes them back to the LLC only on L2 eviction if they are dirty. 

\newcommand\pthreads{pthreads\xspace}

\subsubsection*{Workloads}
\notes{Different from Peacenik, Neat enables optimizations (-O3) for PARSEC programs and default configs for the server programs.}%
Our experiments run the PARSEC 3.0 benchmarks~\cite{parsec-pact-2008}, three real server programs, and the Phoenix benchmarks that contain false sharing~\cite{phoenix}.

For most of the PARSEC benchmarks, we use \code{simsmall} inputs;
we use \code{simmedium} for \bench{swaptions} since \code{simsmall} does not support \textgreater 16 threads.
We use 11 of 13 PARSEC programs; \bench{facesim} fails to finish executing with 
the simulators,
and \bench{freqmine} uses OpenMP instead of \pthreads.
\later{
\brandon{Doesn't OpenMP run with pthreads underneath? freqmine maybe should work...}
}%

The experiments execute three real server programs:
\later{Apache HTTP Server 2.4.23 (\bench{httpd})~\cite{httpd}, Memcached 1.5.2 (\bench{memcached})~\cite{memcached}, and MySQL Server 5.7.16 (\bench{mysqld})~\cite{mysqld}.}%
A\-pa\-che HTTP Server 2.4.23 (\bench{httpd}), Memcached 1.5.2 (\bench{memcached}), and MySQL Server 5.7.16 (\bench{mysqld}).
We configure each program to create a single process 
\notes{(already the default for \bench{mysqld} and \bench{memcached})}%
with 32 worker threads.
For \bench{httpd}, we launch 32 client processes
that repeatedly and concurrently perform simple HTTP requests to HTML pages randomly selected from a pool of 100 pages.
For \bench{memcached} and \bench{mysqld}, we use the benchmark tools \code{memtier\_bench} and \code{sysbench}, respectively, to generate workloads.
Each of the benchmark tools starts 32 client threads to send workloads that perform different mixes of set and get operations (\bench{memcached}) or addition/deletion/update and select operations (\bench{mysqld}).
We configure \code{memtier\_bench} to generate mixed Memcached requests with
the following ratios of gets to sets:
0:100, 10:90, 50:50, and 10:90.
For \bench{mysqld}, \bench{sysbench} generates transactions of mixed SQL queries using the benchmark tool's build-in \emph{read-only (ro)}, \emph{read-write (rw)}, and \emph{write-only (wo)} workloads.
In our experiments, client processes or threads executing natively send 32,768 HTTP requests (\bench{httpd}), 262,144 Memcached requests (\bench{memcached}),
or 8,192 SQL transitions (\bench{mysqld}),
distributed evenly over all 32 client processes or threads.

To evaluate \neat's benefits on false sharing,
our experiments run three Phoenix benchmarks: \bench{histogram}, \bench{linear\_\allowbreak regression}, and \bench{word\_count}.
We selected the benchmarks in which prior work detected false sharing~\cite{huron},
excluding benchmarks in which false sharing is mostly inside pthread functions
(according to Linux's \code{perf c2c} utility), whose code is ignored by our simulators.
As observed by prior work, \code{gcc} eliminates false sharing at certain optimization levels~\cite{predator}.
Our experiments compile each program with \code{gcc} 4.8.5 at the highest optimization level at which false sharing exists:
\bench{histogram} and \bench{linear\_regression} at \code{O1}, and
\bench{word\_count} at \code{O3}.

Our simulators only compute cycles for the ``region of interest'' (ROI).
For each PARSEC or Phoenix program, its ROI includes its whole parallel phase;
\bench{vips} lacks an ROI annotation so we treat its entire execution as the ROI.
For server programs, the ROI is all execution except the startup and shutdown phases.

\subsubsection*{Handling \pthreads functions and atomic instructions}

All simulators identify each \pthreads function call as a synchronization operation.
\later{At each synchronization acquire, all self-invalidation-based \cfgs (\ie, all configurations except \mesi)
perform \si.
Before self-invalidating a dirty line, \neat and \vips commit the line to the LLC.
At each release, the configurations perform different operations to commit dirty data to the LLC as follows.
The \neat \cfgs performs bulk write-backs for dirty lines.
\notes{
The \sarcws \cfg commits to the LLC the write information of a core's current release-free region (RFR) to update other cores' \wss at each release.
}%
The \vips \cfgs commit to the LLC all delayed write-throughs in the write-through buffer.}%
Self-invalidation-based \cfgs
perform relevant coherence actions at synchronization operations,
while MESI ignores all synchronization operations.
All simulators ignore all instructions executed \emph{inside} \pthreads functions.

The \neat, \sarc, and \vips simulators each treat non-\pthreads atomic instructions (\ie,
instructions with the \code{LOCK} prefix) as lock operations but \emph{not} region boundaries,
and thus do not perform any coherence actions but instead execute a distributed queue-based locking protocol~\cite{denovond}.
\Mesi treats atomic instructions as regular memory accesses.

Our \neat (and \sarc and \vips) implementations do not deal with how to implement spin-waiting efficiently under self-invalidation;
we assume a \neat implementation would use an existing mechanism~\cite{callbacks,denovosync,denovond}.

\subsection{Results}

\begin{figure*}[htbp]
  \centering
\vspace*{-2em}
  \hspace*{-1em}
  \subfloat[PARSEC benchmarks]
{\hspace*{-1.2em}\includegraphics[height=1.45in]{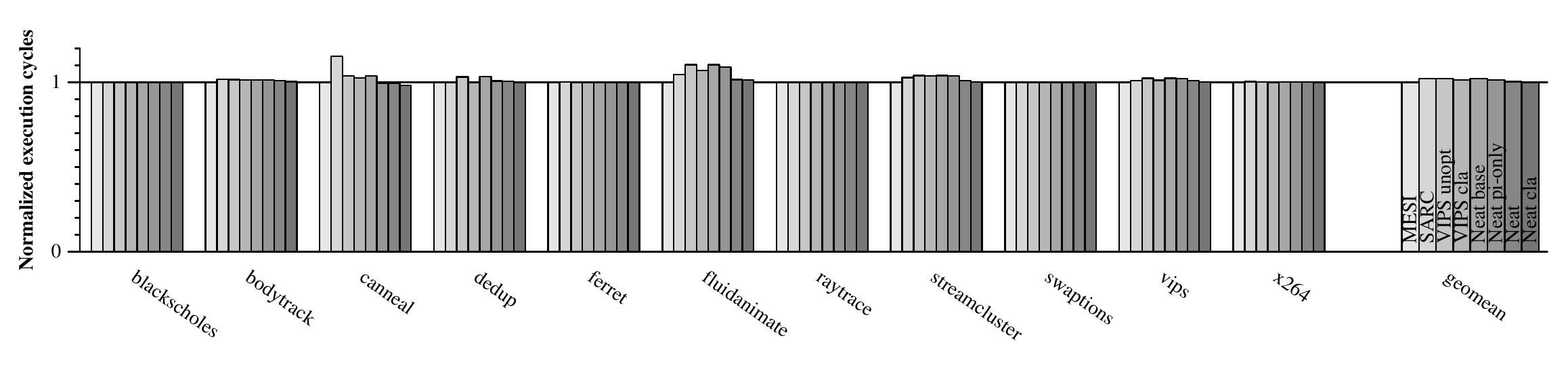}\label{fig:perf}}
  \vspace*{-1.5em}
  \subfloat[Server programs]
{\includegraphics[height=1.45in]{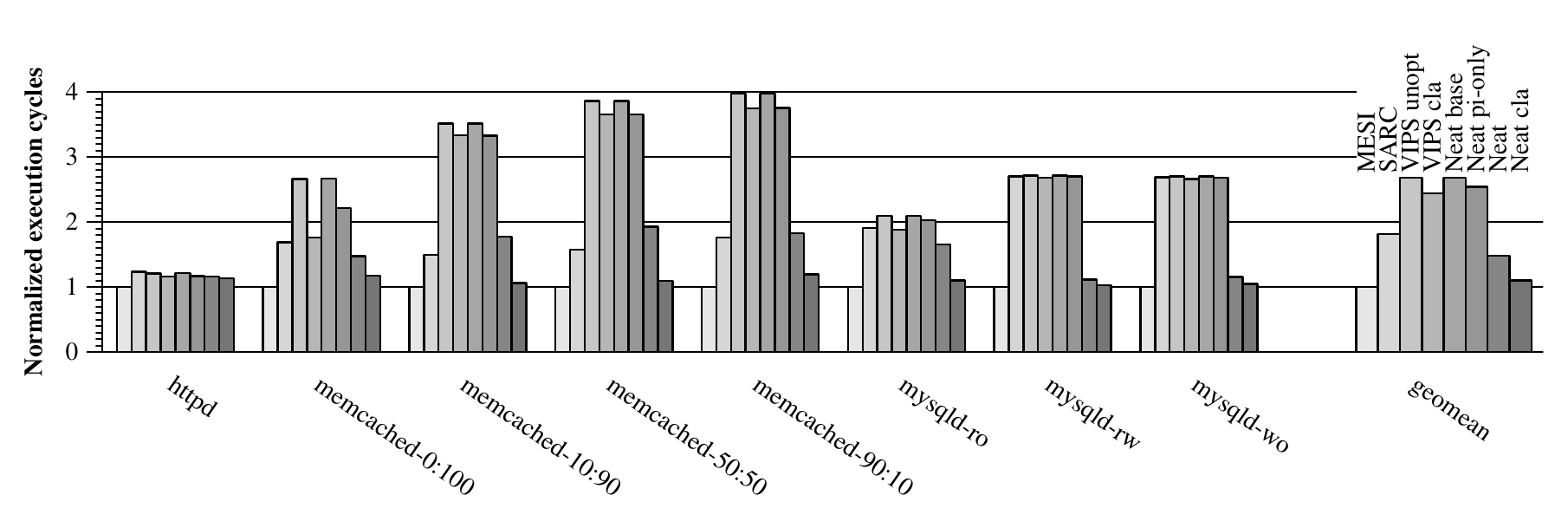}\label{fig:perf-servers}}
  \subfloat[Phoenix benchmarks]
{\includegraphics[height=1.45in]{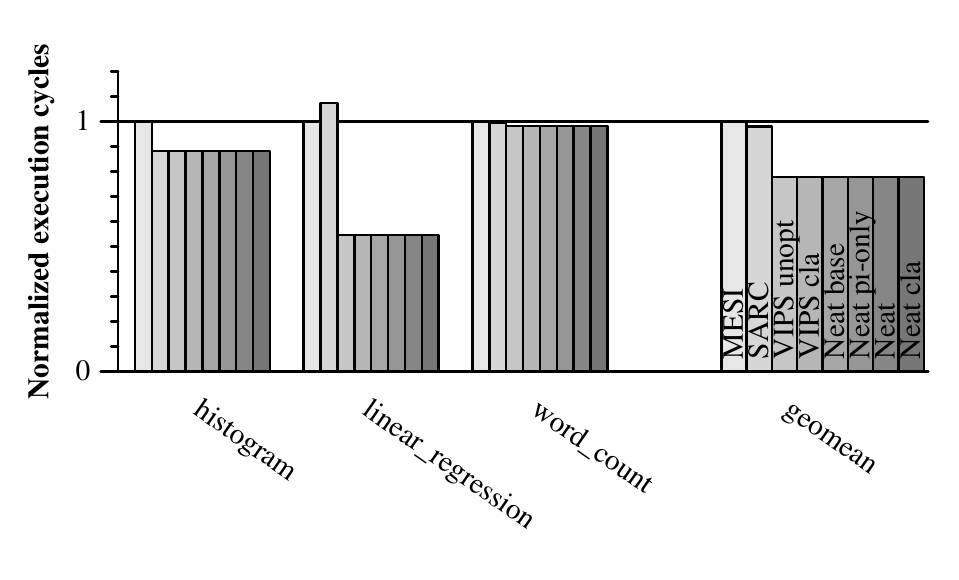}\label{fig:perf-fs}}
  \caption{Execution time for \neat compared with \mesi, \sarc, and \vips on 32 cores, normalized to \mesi.}
  \label{fig:perf-all}
  \centering
\vspace*{-1em}
\hspace*{6em}\includegraphics[width=0.8\linewidth]{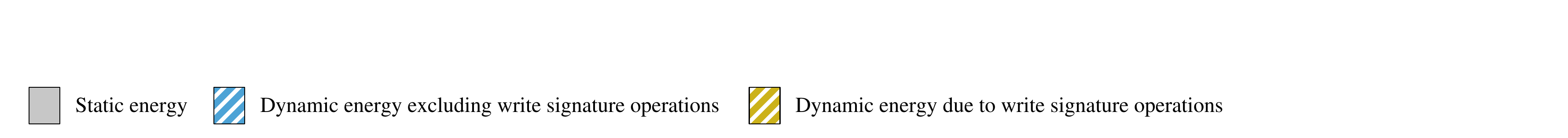}
\vspace*{-2em}\\
  \hspace*{-1em}
  \subfloat[PARSEC benchmarks]
  {\hspace*{-1.2em}\includegraphics[height=1.35in]{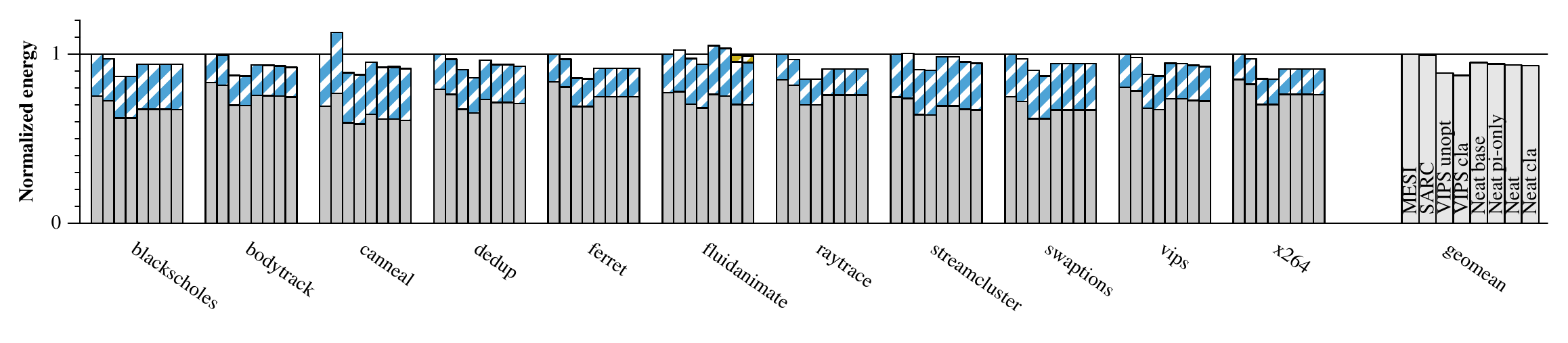}\label{fig:energy}}
  \vspace*{-1.5em}
  \subfloat[Server programs]
  {\includegraphics[height=1.35in]{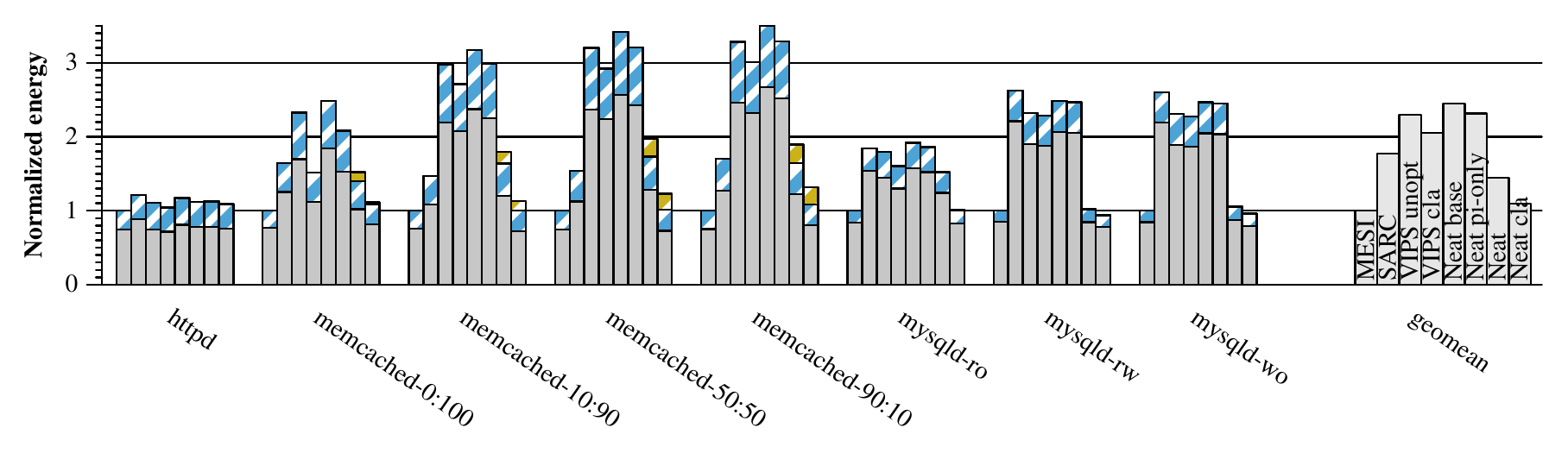}\label{fig:energy-servers}}
  \subfloat[Phoenix benchmarks]
  {\includegraphics[height=1.35in]{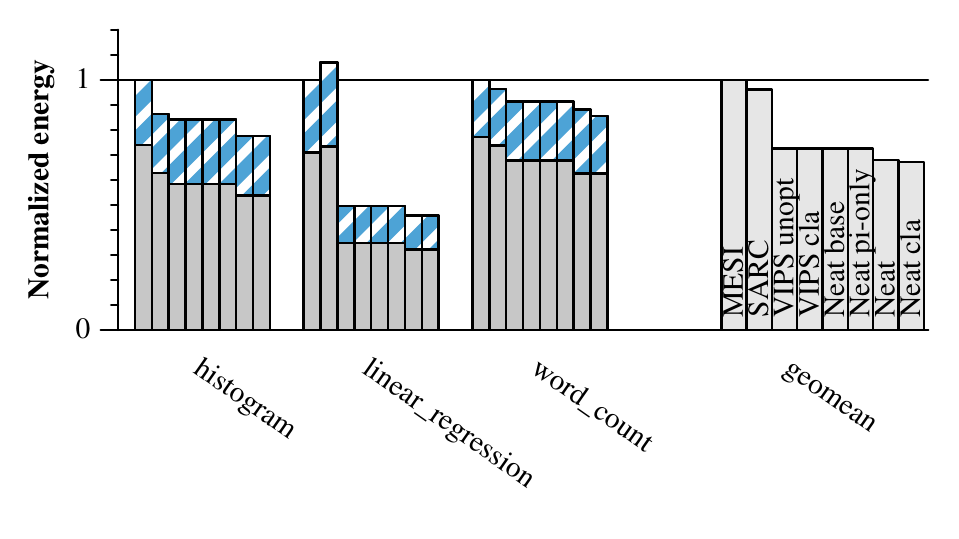}\label{fig:energy-fs}}
  \caption{Energy consumption for \neat compared with \mesi, \sarc, and \vips  on 32 cores, normalized to \mesi.}
  \label{fig:energy-all}
\end{figure*}
\later{\begin{figure*}[t!]
  \centering
  \vspace*{-2.5em}
  \hspace*{-1em}
  \subfloat[PARSEC benchmarks]
  {\hspace*{-1.2em}\includegraphics[height=1.36in]{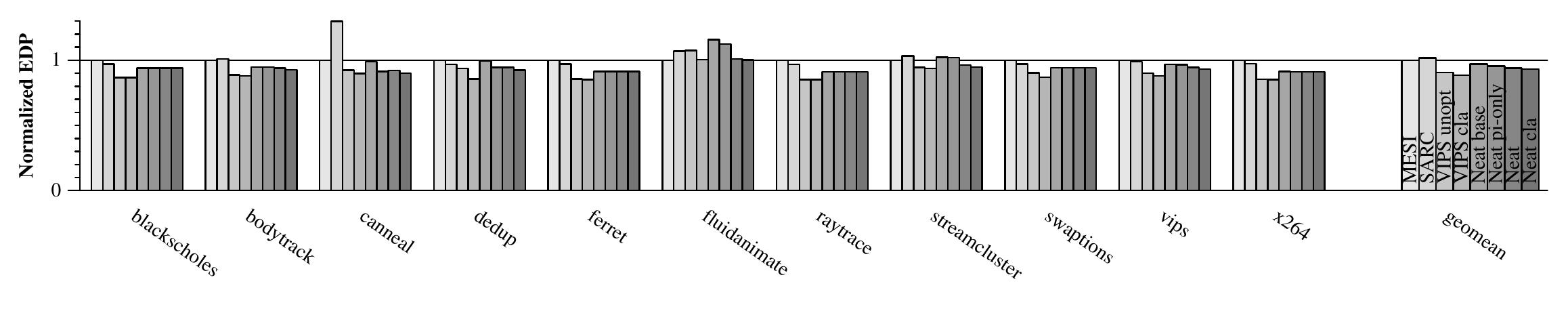}\label{fig:edp}}
  \vspace*{-1.5em}

  \subfloat[Server programs]
  {\includegraphics[height=1.36in]{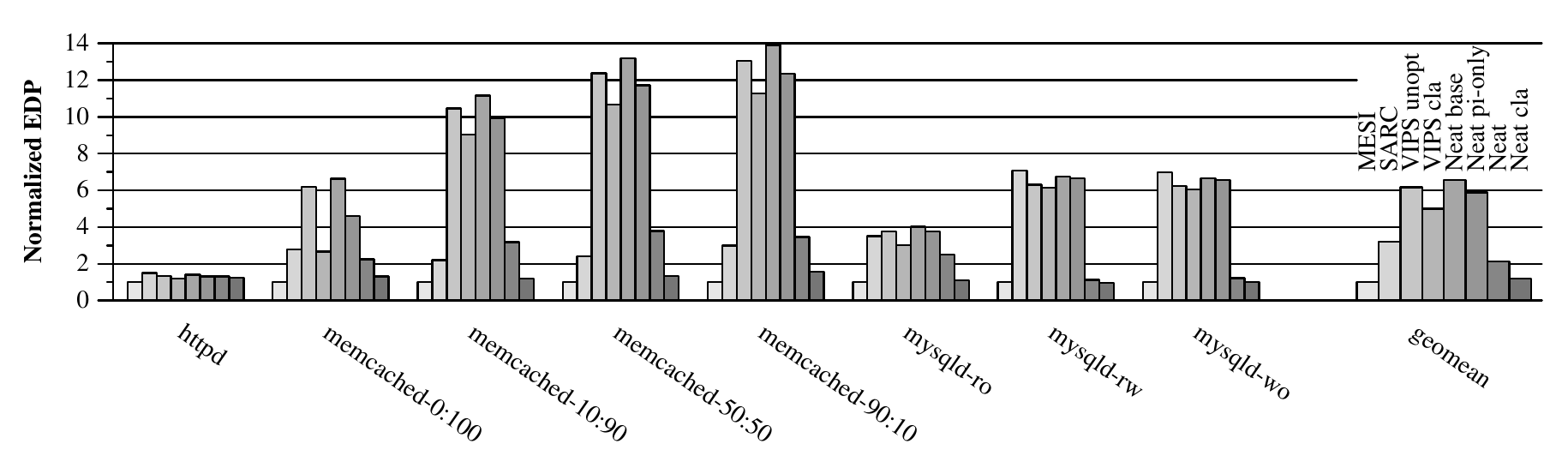}\label{fig:edp-servers}}
  \hfill
  \subfloat[Phoenix benchmarks]
  {\includegraphics[height=1.36in]{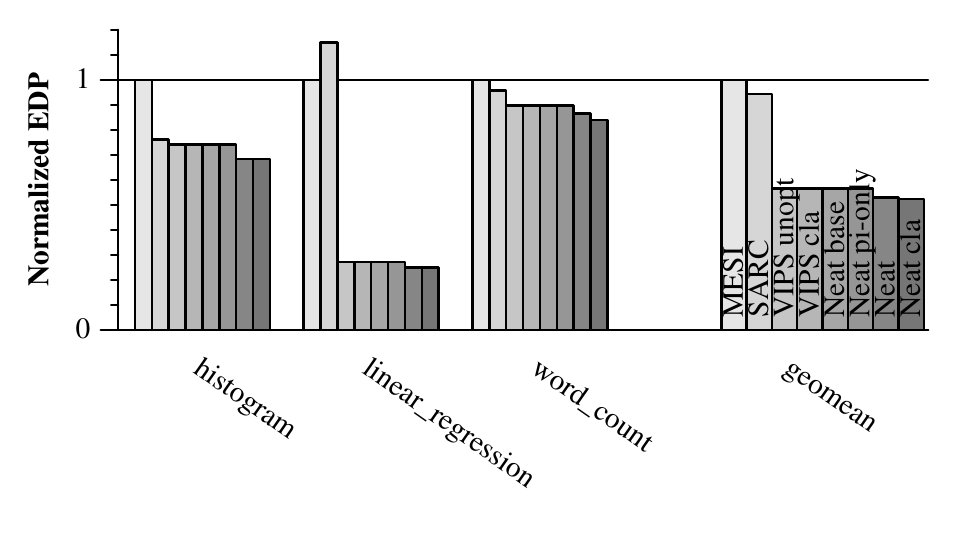}\label{fig:edp-fs}}

  \caption{Energy--delay product (EDP) for \neat compared with \mesi, \sarc, and \vips for all programs on 32 cores, normalized to \mesi.}
  \label{fig:edp-all}
\end{figure*}}%
\later{
\rui{"Why do your results show different performance for VIPS relative to MESI for the same benchmarks?"\\
Not sure how to respond to this concern. Admittedly our simulation model is not as accurate as GEM5 used by VIPS and SARC.
\mike{Hmm, this definitely seems like something to look into.}
\rui{A possible cause of the result discrepancy is that we use different architectural parameters than VIPS, such as private cache levels and sizes (32 KB L1s + 256 KB L2s vs. VIPS's 32 KB L1s). 
The VIPS paper says:"VIPS becomes slightly slower than MESI for large private hierarchies (but less than 3\% for 256KB)"~\cite{vips-directoryless-noc-coherence}.}
\mike{Oh okay, that might be worth looking into.
The VIPS paper also says: ``On average over the 15 applications, VIPS-M is 4.8\% faster
than Directory because of faster writes (no write-misses), faster reads (no directory indirection), and
less traffic in the NoC."}}
}%

Figures~\ref{fig:perf-all} and \ref{fig:energy-all}
\later{ and \ref{fig:edp-all}}%
evaluate \neat's performance and energy consumption,
\later{and power efficiency,}%
respectively, compared with \mesi, \sarc, and \vips, for all programs.
Each bar is normalized to \mesi in all figures.
\later{Figure~\ref{fig:bw-usage} compares the average on-chip bandwidth consumption among the three protocols, without normalization.
    Table~\ref{tab:bench-table} shows how often cores perform synchronization operations, and }%
Table~\ref{tab:stats} shows how many lines are self-invalidated and committed per synchronization operation.
Note that the table's commit counts include (i) all write-throughs made by the \vips configurations at write buffer evictions and synchronization operations and (ii) all write-backs made by the \neat configurations at synchronization operations, but do not include any regular write-backs due to L2 evictions.

\subsubsection*{Performance}

Figure~\ref{fig:perf-all} compares execution cycles (maximum cycles of any core)
for all programs.
The data show that, for most of the PARSEC programs, the \cfgs have similar performance.
For \bench{canneal}, \bench{fluidanimate}, and \bench{streamcluster},
the \sarc and \vips configurations are 3--16\% slower than \mesi.\footnote{%
\renewcommand\sfsmaller{\footnotesize}%
The VIPS paper also evaluated \bench{canneal}, \bench{raytrace}, and \bench{x264}, but reported different results,
presumably due to different architectural parameters used for simulation~\cite{vips-directoryless-noc-coherence}.
The VIPS paper acknowledges that VIPS is slightly (less than 3\%) slower than MESI for 256KB private caches,
which agrees with our results.}
\later{\mike{Comment about the footnote: ``presumably due to different architectural parameters used for simulation''
        sounds like a cop-out, but if we don't know any better than that, it's okay (at least we mentioned it).}}%
In contrast, the full \neat configuration achieves virtually the same performance as \mesi.
On average, \ntopt and \ntcla are nearly at parity with \mesicfg, with an average run-time difference of less than 1\%.

For the server programs, some self-invalidation-based configurations are as much as 4$\times$ slower than \mesi.
\Neat performs significantly better than \sarc and \vips,
with \ntopt
executing 1.2$\times$ and 1.6$\times$ faster than \scunopt and \vipspsrw, respectively.
A run-time breakdown (results not shown) shows that the significant slowdowns for self-invalidation-based configurations
are caused mainly by cache misses due to unnecessary self-invalidation.
Table~\ref{tab:stats} shows that both of \Neat's mechanisms are collectively effective in reducing \sis.
\Vips's classifications are effective, too,
further helping \ntcla achieve performance close to \mesi's
(11\% slower on average).

For two of the Phoenix benchmarks,
\neat and \vips improve performance significantly over \mesi and \sarc
by eliminating writer-initiated invalidations caused by false sharing.
\Neat and \vips do not benefit \bench{word\_count} as much since the program does not contain as much false sharing as the other two.
\Neat's mechanisms and \vips's classifications to avoid unnecessary self-invalidations have negligible impact on performance since the programs all execute few synchronization operations (by using fork--join parallelism).

\subsubsection*{Energy}

Figure~\ref{fig:energy-all} compares the energy consumption of all \cfgs for all the programs, which is divided into energy due to \emph{static} and \emph{dynamic} energy consumption.
For \cfgs with the \ws mechanism, dynamic energy is further divided into energy from \ws operations
and from other sources
(see Section~\ref{sec:eval:perf:impl-meth}).

\Neat imposes a lower static \emph{power} cost than \mesi and \sarc because it eliminates the need to maintain a large, shared coherence directory.
\Neat's mechanisms that avoid unnecessary self-invalidations further help reduce energy \later{, especially for those programs with small regions,}%
since they help reduce \emph{execution time}.
\notes{
($\textnormal{energy} = \textnormal{(static power + dynamic power)} \times \textnormal{time}$).
}%
\Vips imposes an even lower static power than \neat because it does not require per-byte metadata in the private caches as \neat does for the write bits.

On average, for the PARSEC benchmarks, \ntopt (and \ntcla) consume approximately 6\% and 7\% less energy than \mesicfg and \scunopt, respectively, but 6\% more than \vipscla.
For the server programs, \ntopt consumes 44\% more energy than \mesicfg,  
but 18\% and 30\% less than \scunopt and \vipscla, respectively.
\ntcla consumes 9\% more energy than \mesicfg, much less than the \sarc and \vips configurations.
For the Phoenix benchmarks, \ntopt (and \ntcla) consume approximately 32\%, 28\%, and 5\% less than MESI, \scunopt, and \vipscla, respectively.



\later{\subsubsection*{Energy Efficiency}%

To evaluate \barc's energy efficiency, we computed the \emph{energy--delay product} (EDP)~\cite{sarc-coherence} for each \cfg.
Figure~\ref{fig:edp-all} shows the results.
For each PARSEC benchmark, \vips and \neat configurations have similar EDP, and lower EDP than \mesi
(or the same EDP as \mesi for \bench{fluidanimate}).
\ntcla has 7\% lower EDP than \mesi on average.
For server programs, \ntcla has 21\% higher EDP than \mesi but 3$\times$ and 5$\times$ lower EDP than \sarc and \vipscla, respectively.
For Phoenix benchmarks, \ntopt (and \ntcla) has approximately 47\%, 41\%, and 4\% lower EDP than MESI, \scunopt, and \vipscla, respectively.
}

\later{
\subsubsection*{Bandwidth usage}

Figure~\ref{fig:bw-usage} compares the average on-chip bandwidth consumption among \mesi, \sarc, and \neat on 32 cores.
Average on-chip bandwidth consumption is computed as traffic, in 16-byte flit granularity, per unit time.
We define on-chip traffic for \neat as all communication between cores and the LLC.
Traffic for \mesi and \sarc additional includes core-to-core communication.

As shown in Table~\ref{tab:arch-params}, our simulators assume 100 GB/s as the on-chip network's bandwidth limit.
For \bench{canneal}, \neat without the \ws optimization incurs high average on-chip bandwidth usage (90 GB/s), nearly saturating the available on-chip bandwidth.
The fully optimized \ntopt configuration and \sarcws incur comparable on-chip traffic to \mesi for most programs.
}%

\begin{table*}[t]
  \renewcommand\sfsmaller{}
  \centering
  \newcommand{\sicmt}{self-inv/commit\xspace}
  \newcommand{\n}{\ensuremath{n}\xspace}
  \newcommand{\f}{\ensuremath{\mathit{f}}\xspace}
  \newcommand\na{-}
  \renewcommand\k[2]{#2}
  \newcommand\num[1]{0.#1}
  \newcommand\snum[1]{0.0#1}
    \begin{tabular}{@{}l|@{}r@{ / }l|r@{ / }lr@{ / }l|r@{ / }lr@{ / }lr@{ / }lr@{ / }l@{}}
       \bf Program & \multicolumn{2}{c|}{\scunopt} & \multicolumn{2}{c}{\vipsunopt} & \multicolumn{2}{c|}{\vipscla} & \multicolumn{2}{c}{\ntunopt} & \multicolumn{2}{c}{\ntpi} & \multicolumn{2}{c}{\ntopt} & \multicolumn{2}{c@{}}{\ntcla} \\
      \toprule

      \bench{blackscholes} & 142&\na & 64&463 & 25&30 & 64&71 & 2&70 & 0&70 & 0 & 30 \\
      \bench{bodytrack} & 780&\na & 823&304 & 786&48 & 823&82 & 751&41 & 697&41 & 380 & 21 \\

      \bench{canneal} & 3,030&\na & 3,590&182 & 3,420&73 & 3,590&126 & 233&130 & 233&130 & 119 & 54 \\

      \bench{dedup} & 90&\na & 613&3,650 & 82&33 & 613&600 & 123&496 & 95&496 & 12&11 \\

      \bench{ferret} & 159&\na & 335&11,400 & 68&323 & 335&336 & 32&169 & 28&169 & 5&41 \\
      \bench{fluidanimate} & 8&\na & 17&10 & 12&2 & 17&10 & 15&5 & 1&5 & 0&1 \\

      \bench{raytrace} & 563&\na & 589&1,956 & 566&247 & 589&103 & 19&52 & 15&52 & 14&37 \\

      \bench{streamcluster} & 55&\na & 61&4 & 57&2 & 61&4 & 60&4 & 13&4 & 5&2 \\

      \bench{swaptions} & 24&\na & 26&2,640,000 & 10&12 & 26&317 & 3&317 & 0&317 & 0&12 \\

      \bench{vips} & 45&\na & 152&389 & 47&12 & 152&192 & 37&96 & 17&96 & 4&10 \\

      \bench{x264} & 247&\na & 579&10,800 & 341&1,790 & 579&806 & 105&493 & 103&493 & 65&156 \\
     
      \midrule

      \bench{httpd} &  154&\na& 203&235& 166&123& 203&128& 153&68& 144&68& 120&54\\

      \bench{memcached-0:100} &  11 &\na& 25&19& 11&3& 25&20& 22&10& 8&10& 1&2\\
      \bench{memcached-10:90} &  11 &\na& 26&18& 13&4& 26&20& 23&10& 9&10& 1&2\\
      \bench{memcached-50:50} &  9 &\na& 21&15& 12&4& 21&16& 18&8& 6&8& 2&3\\
      \bench{memcached-90:10} &  9 &\na& 19&14& 11&4& 19&15& 16&8& 6&8& 2&3\\

      \bench{mysqld-ro} &  9 &\na& 19&14& 11&4& 19&15& 16&8& 6&8& 2&3\\
      \bench{mysqld-rw} &  102 &\na& 117&23& 104&4& 117&20& 111&10& 9&10& 2&2\\
      \bench{mysqld-wo} &  89 &\na& 103&25& 90&5& 103&20& 96&10& 11&10& 3&3\\

\midrule

\bench{histogram} &  37 &\na& 44&25,341& 26&20,460& 44&51& 2&51& 0&51& 0&46\\
\bench{linear\_regression} &  421,976 &\na& 20&6& 4&6& 20&7& 2&7& 0&7& 0&6\\
\bench{word\_count} &  6,403 &\na& 21&5,516,724& 9&1,513,546& 21&1,399& 2&1,399& 0&1,399& 0&500\\

    \end{tabular}
\vspace*{0.5em}
\caption{Average lines self-invalidated/committed per acquire/release, respectively
  (rounded to 3 significant figures or nearest integer)
  for the PARSEC benchmarks and server programs.
  \textnormal{Note: \Scunopt does not have a commit operation (denoted ``\na'').}}
  \label{tab:stats}
\end{table*}

\later{\begin{table}[t]
  \renewcommand\sfsmaller{}
  \centering
  \newcommand{\n}{\ensuremath{n}\xspace}
  \newcommand{\f}{\ensuremath{\mathit{f}}\xspace}
  \renewcommand\k[2]{#2}
  \newcommand\num[1]{0.#1}
  \newcommand\snum[1]{0.0#1}
  \begin{footnotesize}
    \begin{tabular}{@{}l@{}r@{}}
                   & \bf Avg.\ accesses / \\
       \bf Program & \bf sync.\ operation \\
      \toprule
      \bench{blackscholes} & \k{510}{596,000} \\
      \bench{bodytrack} & \k{85.9}{24,700} \\
      \bench{canneal} & \k{0.009}{35,800} \\
      \bench{dedup} & \k{3.67}{38,600} \\
      \bench{ferret} & \k{103}{209,000} \\
      \bench{fluidanimate}& \k{0.411}{411} \\
      \bench{raytrace} & \k{XXX}{538,000} \\
      \bench{streamcluster} & \k{4.77}{4,860} \\
      \bench{swaptions} & \k{2.34}{20,600,000} \\
      \bench{vips} & \k{15.0}{21,600} \\
      \bench{x264} & \k{45.4}{215,000}
    \end{tabular}~~~\vrule~~~%
    \begin{tabular}{@{}l@{}r@{}}
                   & \bf Avg.\ accesses / \\
       \bf Program & \bf sync.\ operation \\
      \toprule
      \bench{httpd} & \k{0.345}{2,330} \\
      \bench{memcached-0:100} & \k{0.037}{92} \\
      \bench{memcached-10:90} & \k{0.037}{85} \\
      \bench{memcached-50:50} & \k{0.037}{59} \\
      \bench{memcached-90:10} & \k{0.037}{47} \\
      \bench{mysqld-ro} & \k{0.058}{435} \\
      \bench{mysqld-rw} & \k{0.058}{255} \\
      \bench{mysqld-wo} & \k{0.058}{264} \\\\\\\\
    \end{tabular}
  \end{footnotesize}
  
  \caption{Average memory accesses per synchronization operation
  (rounded to 3 significant figures or nearest integer)
  for the PARSEC benchmarks and server programs.}
  \label{tab:bench-table}
\end{table}}%

\medskip
\noindent
In summary, the results in this section show that for less-complex programs such as the PARSEC benchmarks,
\neat has performance and energy competitive
\later{and on-chip bandwidth consumption}%
with \mesi and outperforms \sarc and \vips.
\Neat has significant benefits over \mesi and \sarc specifically for programs with false sharing.
For complex server programs, \neat is slower and consumes more energy than \mesi, but still has significant benefits over \sarc and \vips.
\notes{
\Neat's \ws \opt is applicable to \sarc and improves \sarc to be comparable to \neat.
Similarly,
}%
\Vips's classifications benefit \neat and improve \neat
to be competitive with \mesi.
The evaluations in this section and Section~\ref{sec:protocol-verification} show that \neat is not only a novel system design,
but it improves in complexity compared to \mesi and improves in performance and energy consumption compared to \sarc and \vips (and \mesi for programs with false sharing).

\section{Related Work}
\label{Sec:related}


Sections~\ref{sec:background} compared \neat
qualitatively with closely related work that uses self-invalidations~\cite{sarc-coherence,vips-directoryless-noc-coherence, denovo, denovond, denovo-gpu,hlrc,quickrelease}.
This section discusses other related work.

\subsubsection*{Other DeNovo-based work}

\emph{DeNovoSync} uses DeNovo's protocol that employs \si and registration,
but---like SARC, VIPS, and \barc---DeNovoSync applies to general shared-memory programs~\cite{denovosync}.
\later{Since in a setting without writer-initiated invalidations,
that is widely used in traditional coherence protocols (\eg, MESI)}%
DeNovoSync's contribution focuses on supporting arbitrary synchronization operations in the context of \si.
\later{by registering synchronization reads for ownership}%
The evaluation limits self-invalidation costs by assuming programmer annotations.
In contrast,
\later{the baseline \barc conservatively self-invalidates all touched data at acquire operations and the optimized}%
\barc uses automatic mechanisms to reduce self-invalidation
and assumes no additional knowledge.




\emph{Spandex} provides a flexible interface that supports various coherence protocols, including \mesi, GPU coherence, and DeNovo~\cite{spandex}.
It necessarily suffers MESI's complexity overhead, including transient states, Inv and Ack messages, and support for core-to-core communication.
\later{Similar to DeNovo, the Spandex LLC registers written data to track ownership and thus requires inclusivity.}%
In contrast, \neat does not need to support MESI or other protocols.
\later{It does not have the complexity overhead of Spandex, including transient states in the Spandex LLC and support for core-to-core communication.
    In fact, by designing \neat, we argue that CPUs can use simple GPU-style protocols and thus it is not necessary to have a complex coherence interface like Spandex that aims to support both complex MESI-style CPU protocols and simple GPU protocols.
Further, \neat does not require an inclusive LLC since it does not track ownership in the LLC.}%

\later{
\mike{What's \cite{dsi-isca-95} about?}
}%

\subsubsection*{Exploiting self-invalidation-based coherence}

\emph{ARC} uses self-invalidation and defers dirty write-backs until release operations~\cite{arc}.
While \neat assumes data race freedom (DRF),
ARC \emph{checks} DRF,
leading to a more complex system.
Although ARC optimizes self-invalidation and dirty write-back costs,
its optimizations differ from \neat's by leveraging mechanisms for performing conflict detection.

Jimborean \etal\ use compile-time analysis to detect \emph{extended DRF regions}
and thus reduce the frequency of self-invalidation~\cite{extended-drf-regions}.
Extended DRF regions would apply to \barc.
\later{
Kaxiras and Ros show how to apply self-invalidation mechanisms to simplify and optimize
virtually addressed caches and bus coherence~\cite{vips-virtual-cache-coherence,vips-snoopless-bus-coherence}.}%

\subsubsection*{Using write signatures to represent write sets}

Prior work has used write signatures to represent a core's own write set~\cite{denovond,tc-release-pp}. In contrast, in \neat a core's write signature represents \emph{writes by all other cores} (and the signatures are thus maintained at the LLC).
An exception is \emph{Racer}~\cite{racer-tso}, which maintains write signatures in a similar way to \neat, but for a distinct purpose: detecting read-after-write races in order to treat them as synchronization points.

\section{Conclusion}


\Neat is a new cache coherence design that avoids unnecessary self-invalidations
and performs bulk write-backs at synchronization operations.
Unlike other self-invalidation approaches,
\neat does not rely on programmer annotations or
specific access patterns for efficiency.
Our evaluation shows that \neat is simpler than MESI,
performs competitively with MESI especially under false sharing,
and provides better performance and energy efficiency than two state-of-the-art self-invalidation-based
approaches.
These results suggest that
\Neat provides efficient, complexity-effective coherence.

\section*{Acknowledgments}

We thank Rakesh Komuravelli \etal\ for sharing the Murphi specification of MESI
from their work~\cite{denovo-verification}.
This material is based upon work supported by the National Science Foundation
under Grants CAREER-1253703, CCF-1421612, XPS-1629126, and XPS-1629196.

\bibliographystyle{abbrv}
\bibliography{bib/conf-abbrv,bib/plass}

%
%

\end{document}